\documentclass{article}
\usepackage[textwidth=1.2\textwidth]{geometry}
\usepackage[utf8]{inputenc}
\usepackage[T1]{fontenc}
\usepackage[final]{fixme}
\usepackage{array}
\usepackage{comment}
\usepackage{randtext}
\usepackage{stmaryrd} 
\usepackage{upgreek}
\usepackage{interval}
\usepackage{enumitem}
\usepackage{amsmath}
\usepackage{amssymb}
\usepackage{eqntabular,renumber,galois,calculus}
\usepackage{graphicx}
\usepackage{relsize}
\usepackage{phaistos}
\usepackage{multicol}
\usepackage{placeins}
\usepackage[british]{babel}
\usepackage{tikz}
\usetikzlibrary{decorations.pathreplacing}
\usepackage{hierarchy}
\usepackage{invariance}

\newcommand{\catbeginrmw}{\hyperlink{hyper:catbeginrmw}{\texttt{beginrmw}}\xspace}

\newcommand{\catendrmw}{\hyperlink{hyper:catendrmw}{\texttt{endrmw}}\xspace}
\usepackage{interval}

\usepackage{enumitem}
\usepackage{xspace}
\makeatletter
\newcommand{\ie}{\textit{i.e}\@ifnextchar.{}{.\xspace}}
\newcommand{\eg}{\textit{e.g}\@ifnextchar.{}{.\xspace}}
\newcommand{\etc}{\textit{etc}\@ifnextchar.{}{.\xspace}}
\newcommand{\ao}{\textit{a.o}\@ifnextchar.{}{.\xspace}}
\newcommand{\QED}{\textit{Q.E.D}\@ifnextchar.{}{.\xspace}}
\newcommand{\issuchthat}{\textit{s.t}\@ifnextchar.{}{.\xspace}}
\newcommand{\viz}{\textit{viz}\@ifnextchar.{}{.\xspace}}
\newcommand{\marge}{\hskip\parindent\phantom{\labelitemi}\hskip\labelsep}
\newcommand{\nbsp}{{\mbox{$~$}}}
\newif\ifitalicenv\italicenvtrue

\newtheorem{theorem}{Theorem}[section]

\newtheorem{lemma}[theorem]{Lemma}

\newtheorem{intuition}[theorem]{Intuition}

\newtheorem{exam}[theorem]{Example}
\newenvironment{example}{\italicenvfalse
\begin{exam}}{\end{exam}\italicenvtrue}
\newtheorem{defi}[theorem]{Definition}

\def\@begintheorem#1#2{    \trivlist
    \item[        \hskip 12\p@
        \hskip \labelsep
        {\ifitalicenv\sc\else\itshape\fi #1\hskip 5\p@\relax{\rm #2}.\enspace}]        \ifitalicenv\itshape\else\upshape\fi\hskip-\labelsep}
\def\@opargbegintheorem#1#2#3{    \trivlist
    \item[\hskip 12pt
          \hskip \labelsep
          {\ifitalicenv{\sc{#1}}\else{\itshape#1}\fi	   \savebox\@tempboxa{\ifitalicenv{\scshape#3}\else{\itshape#3}\fi}	   \ifdim\wd\@tempboxa>\z@           \ {\rm #2}\unskip\hskip5pt\relax$($\box\@tempboxa$)$	   \fi.\unskip\hskip5pt}]
\ifitalicenv\itshape\else\upshape\fi\hskip-\labelsep}

\newif\if@qeded
\global\@qededfalse
\def\proof{    \global\@qededfalse
    \@ifnextchar[{\@xproof}{\@proof}}

\def\endproof{    \if@qeded\else\qed\fi
    \endtrivlist
}
\def\@proof{    \trivlist
    \item[        \hskip 12\p@
        \hskip \labelsep
        {\sc Proof.\enspace}]\hskip-\labelsep    \ignorespaces
}
\def\@xproof[#1]{    \trivlist
    \item[\hskip 12\p@\hskip \labelsep{\sc Proof #1.}]    \ignorespaces
}
\def\qed{\unskip\kern 10pt{\unitlength1pt\linethickness{.4pt}\framebox(5,5){}}
    \global\@qededtrue
    }\def\newdef#1#2{    \expandafter\@ifdefinable\csname #1\endcsname
        {\@definecounter{#1}         \expandafter\xdef\csname the#1\endcsname{\@thmcounter{#1}}         \global\@namedef{#1}{\@defthm{#1}{#2}}         \global\@namedef{end#1}{\@endtheorem}    }}
\def\@defthm#1#2{    \refstepcounter{#1}    \@ifnextchar[{\@ydefthm{#1}{#2}}{\@xdefthm{#1}{#2}}}
\def\@xdefthm#1#2{    \@begindef{#2}{\csname the#1\endcsname}    \ignorespaces
}
\def\@ydefthm#1#2[#3]{    \trivlist
    \item[        \hskip 10\p@
        \hskip \labelsep
        {\it #2         \savebox\@tempboxa{#3}         \ifdim \wd\@tempboxa>\z@
            \ \box\@tempboxa
         \fi.        }]    \ignorespaces
}
\def\@begindef#1#2{    \trivlist
    \item[        \hskip 10\p@
        \hskip \labelsep
        {\it #1\ \rm #2.}    ]}

\makeatother

\usepackage{hyperref}

\title{\textbf{\LARGE Syntax and analytic semantics of LISA}}
\author{\begin{tabular}{c}
\Large\bfseries Jade Alglave \\[1ex]
Microsoft Research Cambridge \\ 
University College London\\
\texttt{\randomize{jaalglav@microsoft.com}}, \texttt{\randomize{j.alglave@ucl.ac.uk}}\\[1em]
\Large\bfseries Patrick Cousot\\[1ex]
New York University\\ 
emer. \'Ecole Normale Sup\'erieure, PSL Research University\\
\texttt{\randomize{pcousot@cims.nyu.edu}}, \texttt{\randomize{cousot@ens.fr}}\\[1ex]
\end{tabular}}

\usepackage{natbib}
\def\cite{\citep*}
\begin{document}

\newif\ifInvariance
\Invariancefalse

\maketitle

\newcommand{\refsection}[1]{Sect.\nbsp\ref{#1}}
\newcommand{\reffigure}[1]{Fig.\nbsp\ref{#1}}

\begin{abstract}
We provide the syntax and semantics of the \LISA (for ``Litmus Instruction
Set Architecture'') language. The parallel assembly language \LISA is implemented in the \href{http://virginia.cs.ucl.ac.uk/herd/}{\texttt{herd7}} tool \cite{herd7} for
simulating weak consistency models. 
\end{abstract}

\section{Introduction}

\LISA (which stands for ``Litmus Instruction
Set Architecture'') has the vocation of being a fairly minimal assembly
language, with read and write memory accesses, branches and fences to design consistency models for weakly consistent systems  without having to
concern oneself with the syntax of the programming language (such as ARM, IBM, Intel x86, Nvidia multiprocessor chips, or languages like C++ or
OpenCL), which has proved
quite useful at times where said syntax was still in flux. 

The  weakly consistent semantics of a \LISA is analytic in that it is the intersection of an anarchic semantics (without any restriction on communications) and a communication semantics (specified by a {\tt cat} specification \cite{AlglaveCousotMaranget-cat-HSA-2015} restricting the allowed communications).

The \href{http://virginia.cs.ucl.ac.uk/herd/}{\texttt{herd7}} tool is a weakly consistent system 
simulator, which takes as input a {\tt cat} specification \cite{AlglaveCousotMaranget-cat-HSA-2015} and a litmus test preferably in \LISA,
and determines whether the candidate executions of this test are allowed or not
under the {\tt cat} specification and under which conditions on communication events.  The semantics of {\tt
cat} and \LISA has been implemented in the \href{http://virginia.cs.ucl.ac.uk/herd/}{\texttt{herd7}} tool. The documentation of the tool is available
online, at
\href{http://diy.inria.fr/tst7/doc/herd.html}{\url{diy.inria.fr/tst7/doc/herd.html}}.
The sources of the tool are available at \href{http://diy.inria.fr}{\url{diy.inria.fr}}. A web
interface of \href{http://virginia.cs.ucl.ac.uk/herd/}{\texttt{herd7}} is available at
\href{http://virginia.cs.ucl.ac.uk/herd/}{\url{virginia.cs.ucl.ac.uk/herd}}.

We define the anarchic true parallel semantics with separated communications of \LISA where anarchic means that no restriction is made on possible communications.
We also formally define the abstraction into candidates executions which are the inputs for the semantics of the \cat\ language placing restriction on the communication events, which defines a weak consistency model.

\section{An overview of analytic semantics}

We introduce anarchic semantics with true parallelism and unrestricted separate communications in Section \ref{sec:Anarchic-semantics}. Then in Section \ref{sec:cat-specification-weakly-consistent-semantics} we show how to abstract anarchic execution to candidate executions that a \texttt{cat} specification will allow or forbid based on hypotheses on relations between communication events.

\begingroup
\let\subsubsection\subsection
\def\subsection#1{}
\newcommand{\setofallpythiavariablesoperator}[1]{\mathfrak{P}\def\@setofallpythiavariables{#1}\ifx\@setofallpythiavariables\@empty\else(#1)\fi}
\newcommand{\defsetofallpythiavariables}[1]{\hypertarget{hyper:setofallpythiavariables}{\setofallpythiavariablesoperator{#1}}}
\newcommand{\setofallpythiavariables}[1]{\hyperlink{hyper:setofallpythiavariables}{\setofallpythiavariablesoperator{#1}}}
\subsection{Anarchic semantics}\label{sec:Anarchic-semantics}

\subsubsection{Executions} The anarchic semantics of a parallel program is a set of
executions; an execution has the form $\execution$ =
$\computation\times\rfcommunicationrelation\in\defsetofallexecutions$, where $\computation$ is the
\emph{computation} part and $\rfcommunicationrelation$ is the
\emph{communication} part. 

\paragraph{Communications $\!\!\!\!\!\!$} are sets $\rfcommunicationrelation$,
which gather \emph{read-from} relations. A read-from relation $\catrfevent{w}{r}$
links a (possibly initial) write event $w$ and a read event $r$ relative to the same shared
variable $\avar{x}$ with the same value. Communications are \emph{anarchic}: we
place no restriction on which write a read can read from; restrictions can be
made in a \cat specification however.

\paragraph{Computations $\!\!\!\!\!\!$} have the form $\computation$ =
$\parallelexecution{\trace_{\startprocess}}{p\in\setofallprocessidentifiers}{}{\trace_p}{}{}$, where
$\trace_{\startprocess}$ is an execution \emph{trace} of the prelude process,
and $\trace_p$ are execution traces of the processes
$p\in\setofallprocessidentifiers$.
A finite (resp. infinite) non-empty trace $\trace_p$,
$p\in\setofallprocessidentifiersstart$ is a finite (resp. infinite)
sequence 
\bgroup\ifInvariance
\jot=0pt\abovedisplayskip-3pt\belowdisplayskip2pt
\fi
\begin{eqntabular*}{rclcl}
\trace_p & = &\langle\eventstate{\eventsoftrace{\trace_p}_{k}}{\statesoftrace{\trace_p}_{k}}\mid
k\in\interval[open right]{0}{1+m}\rangle&\in&\defsetofalltraces
\end{eqntabular*}\egroup 
of computation steps $\ulstrut\smash{\eventstate{\eventsoftrace{\trace_p}_{k}}{\statesoftrace{\trace_p}_{k}}}$
(with ${\eventsoftrace{\trace_p}_{k}}$ an \emph{event} and ${\statesoftrace{\trace_p}_{k}}$ the next \emph{state}---see below
for the definitions of event and state)
such that ${\eventsoftrace{\trace_p}_{0}}=\defstartevent$ is the start event,
$\lengthoftrace{\trace_p}\triangleq m\in\setofallstrictlypositive$ for finite traces and $\lengthoftrace{\trace_p}\triangleq m=\omega=1+\omega$ for
infinite traces where $\omega$ is the first infinite limit ordinal so that $\interval[open right]{0}{1+\omega}=\interval[open right]{0}{\omega}=\setofallnaturals$. This is a true parallelism formalisation since there is a
notion of local time in each trace $\trace_p$,
$p\in\setofallprocessidentifiersstart$ of an execution $\execution$ =
$\parallelexecution{\trace_{\startprocess}}{p\in\setofallprocessidentifiers}{}{\trace_p}{}{\rfcommunicationrelation}$
but no global time, since it is impossible to state that an event of a process
happens before or after an event of another process or when communications do
happen.

\paragraph{Events $\!\!\!\!\!\!$} indicate several things:
\begin{itemize}[align=left,leftmargin=*,topsep=0pt,itemsep=0pt,parsep=0pt]
\item their \emph{nature}, \eg{} read (\ttfont{r}), write (\ttfont{w}), branch (\ttfont{b}), fence (\ttfont{f}), \etc;
\item the \emph{identifier} $p$ of the process that they come from;
\item the \emph{control label} $\ell$ of the instruction that they come from;
\item the \emph{instruction} that they come from---which gives the shared
variables and local registers affected by the event, if any, \eg{} {\tt x} and
{\tt R} in the case of a read
$\rinstra[\tagsequence]{{\areg{R}}}{\avar{x}}{}$;
\item their \emph{stamp} $\theta\in\defsetofallstamps{p}$; they ensure that events in a trace are
unique. In our examples, stamps gather the control label and iteration counters
of all surrounding loops, but this is not mandatory: all we need is for events
to be uniquely stamped. Different processes have non-comparable stamps. Stamps are
totally ordered per process by $\mathord{\defltstamp{p}}$ (which is irreflexive and transitive, while events on different processes are different and incomparable).  The successor
function $\defsuccstamp{p}$ is \issuchthat
$\theta\ltstamp{p}\succstamp{p}(\theta)$ (but not necessarily the immediate
successor); $\definfstamp{p}$ is a minimal stamp for process $p$. We consider
executions  up to the isomorphic order-preserving renaming  $\traceequivalence$ of stamps;
\item their \emph{value} $v\in\defsetofallvalues$, whether ground or symbolic. To name the values that
are communicated in invariants, we use pythia variables $\defsetofallpythiavariables{p}\triangleq\{ \svar{x}{\theta}
\mid \avar{x}\in\setofallavariables\wedge \theta\in\setofallstamps{p}\}$ (note
that the uniqueness of stamps on traces ensures the uniqueness of pythia
variables). More precisely, traditional methods such as Lamport's and
Owicki-Gries' name \avar{x} the value of the shared variable \avar{x}, but we
cannot use the same idea in the context of weak consistency models. Instead we
name $\svar{x}{\theta}$ the value of shared variable \avar{x} read at local
time $\theta$.  \end{itemize}

The events $\eventsoftrace{\tau_{p}}$ on a trace $\tau_{p}$ of process $p$ are
as follows\makeatletter\@ifundefined{r@fig:com-scenar}{}{ (\eg \reffigure{fig:com-scenar})}\makeatother:
\begin{itemize}[align=left,leftmargin=*,topsep=0pt,itemsep=0pt,parsep=0pt]

\item register events:
$\assignmentevent{\markertuple{p}{\ell}{\mova{{\areg{R}_1}}{\nonterminal{operation}}}{\theta}}{v}$;

\item read events:
${\readevent{\markertuple{p}{\ell}{\rinstra[\tagsequence]{{\areg{R}_1}}{\avar{x}}{}}{\theta}}{\svar{x}{\theta}}})$;

\item write events:
${\writeevent{\markertuple{p}{\ell}{\winstra[\tagsequence]{\avar{x}}{}{\nonterminal{r-value}}}{\theta}}{v}}$;

\item branch events are of two kinds:
  \begin{itemize}[align=left,leftmargin=*,topsep=0pt,itemsep=0.pt,parsep=0pt]
  \item
$\testevent{\markertuple{p}{\ell}{\binstra[\tagsequence]{\nonterminal{operation}}{\grlabel_{\mskip-2mu
t}}{}}{\theta}}{}$ for the true branch;
  \item $\testevent{\markertuple{p}{\ell}{
\binstra[\tagsequence]{\mneg\nonterminal{operation}}{\grlabel_{\mskip-2mu
t}}}{\theta}}{}$ for the false branch;
  \end{itemize}

\item fence events: $\markerevent{\markertuple{p}{\ell}{\finstra{\tagsequence}\
\bigl[\{\grlabel_1^0 \ldots \grlabel_1^m\}\ \{\grlabel_2^0 \ldots
\grlabel_2^q\}\bigr]}{\theta}}$;

\item RMW events are of two kinds: 
  \begin{itemize}[align=left,leftmargin=*,topsep=0pt,itemsep=0pt,parsep=0pt]
  \item begin event: $\markerevent{\markertuple{p}{\ell}{\grbeginrmw[\tagsequence]{x}}{\theta}}{}$;
  \item end event: ${\markerevent{\markertuple{p}{\ell\allowbreak}{\grendrmw[\tagsequence]{x}}{\theta}}}$
  \end{itemize}
\end{itemize}

\paragraph{States $\!\!\!\!\!\!$}
$\sigma=\defstateofprocess{\ell}{\theta}{\aenvironment}{\avaluation}$ of a
process $p$ mention:
\begin{itemize}[align=left,leftmargin=*,topsep=0pt,itemsep=0.pt,parsep=0pt]
\item $\ell$, the current control label of process $p$ (we have
$\attdone[\texttt{P}](p)\ell$ which is $\true$ if and only if $\ell$ is the
last label of process $p$ which is reached if and only if process $p$ does
terminate);
\item $\theta$ is the stamp of the state in process $p$; 
\item $\aenvironment$ is an environment mapping the local registers $\areg{R}$
of process $p$ to their ground or symbolic value $\aenvironment(\areg{R})$; 
\item $\avaluation$ is a valuation mapping the pythia variables $\svar{x}{\theta}\in\setofallpythiavariables{p}$  of a process $p$ to
their ground or symbolic value $\avaluation(\svar{x}{\theta})$. This is a partial map since the pythia
variables (\ie{} the domain $\domain(\avaluation)$ of the valuation
$\avaluation$) augment as communications unravel. Values can be ground, or
symbolic expressions over pythia variables.  
\end{itemize}
The prelude process has no state (represented by $\bullet$).

\subsubsection{Well-formedness conditions}

We specify our anarchic semantics by the means of well-formedness conditions over the
computation traces
$\computation=\parallelexecution{\trace_{\startprocess}}{p\in\setofallprocessidentifiers}{}{\trace_p}{}{}$,
and the communications $\rfcommunicationrelation$ of an execution
$\execution=\computation\times\rfcommunicationrelation$. 

\paragraph{Conditions over computations $\!\!\!\!\!\!$}
$\parallelexecution{\trace_{\startprocess}}{p\in\setofallprocessidentifiers}{}{\trace_p}{}{}$
are as follows:

\begin{itemize}[align=left,leftmargin=*,topsep=0pt,itemsep=2pt,parsep=0pt]

\item \textit{Start}: traces $\tau$ must all start with a unique fake start event $\startevent$:
\bgroup\abovedisplayskip0.5\abovedisplayskip\belowdisplayskip0.5\belowdisplayskip\begin{eqntabular}[fl]{@{}l@{}}
\eventsoftrace{\tau}_{0}=\startevent\wedge\forall i\in\interval[open left, open right]{0}{1+\lengthoftrace{\tau}}\suchthat\eventsoftrace{\tau}_{i}\neq\startevent\ .
\defcondition<start>{start}{\execution}\end{eqntabular}\egroup

\item \textit{Uniqueness}: the stamps of events must be unique:
\bgroup\abovedisplayskip0.5\abovedisplayskip\belowdisplayskip0.5\belowdisplayskip\begin{eqntabular}[fl]{@{}l@{}}
\forall p,q\in\setofallprocessidentifiers\cup\{\startprocess\}\suchthat
\forall i\in\interval[open right]{0}{1+\lengthoftrace{\tau_p}}\suchthat
\forall j\in\interval[open right]{0}{1+\lengthoftrace{\tau_q}}\suchthat\nonumber\\
(p=q\implies i\neq j)\implies{}\stampof(\eventsoftrace{\tau_p}_i)\neq\stampof(\eventsoftrace{\tau_q}_j)\ .
\defcondition<uniqueness>{uniqueness}{\execution}
\end{eqntabular}\egroup
It immediately follows that events of a trace are unique, and the pythia
variable $\svar{x}{\theta}$ in any read
$\readevent{\markertuple{p}{\ell}{\areg{r} \aassign
\avar{x}}{\theta}}{\svar{x}{\theta}}$ is unique.

\item \textit{Initialisation}: all shared variables $\avar{x}$ are initialised
once and only once to a value $v_{\avar{x}}$ in the prelude (or to
$v_{\avar{x}}=0$ by default).
\bgroup\abovedisplayskip0.5\abovedisplayskip\belowdisplayskip0.5\belowdisplayskip\begin{eqntabular}[fl]{@{}l}
\exists \tau_0,\tau_1\suchthat {\trace_{\startprocess}} = \tau_0 \:(\writeevent{\markertuple{\startprocess}{\defstartlabel}{\avar{x} \aassign e}{\theta}}{v_{\avar{x}}})\: \tau_1
\wedge{}\nonumber\\
\forall \tau_0,\tau_1,\tau_2\suchthat ({\trace_{\startprocess}} = \tau_0\smash{\eventstate{\epsilon}\,{\sigma}} \tau_1 \, \writeevent{\markertuple{\startprocess}{\ell}{\avar{x} \aassign e}{\theta}}{v_{\avar{x}}}\: \tau_2)\nonumber\\
\quad{}\implies{}(\epsilon\in\setofallwriteevents(\startprocess,\avar{y})\wedge \avar{y}\neq\avar{x})\ .
\defcondition<initialisation>{initialisation}{\execution}
\end{eqntabular}\egroup
\item \textit{Maximality}: a finite trace $\tau_p$ of a process $p$ must be
maximal \ie{} must describe a process whose execution is finished. Note that
infinite traces are maximal by definition, hence need not be included
in the following maximality condition: 
\bgroup\abovedisplayskip0.5\abovedisplayskip\belowdisplayskip0.5\belowdisplayskip\begin{eqntabular}[fl]{@{}l@{}}
\exists \ \ell \ \theta \ \aenvironment \ \avaluation\suchthat
{\statesoftrace{\tau_p}}_{\lengthoftrace{\tau}}=\stateofprocess{\ell}{\theta}{\aenvironment}{\avaluation}
\wedge \attdone[\texttt{P}](p)\ell\ \defcondition{maximality}{\execution}
\end{eqntabular}\egroup
\ie the control state of the last state of the
trace is at the end of the process, as indicated by
$\attdone[\texttt{P}](p)\ell$.  \end{itemize}

\paragraph{Conditions over the communications $\!\!\!\!\!\!$}
$\rfcommunicationrelation$ are as follows: 

\begin{itemize}[align=left,leftmargin=*,topsep=0pt,itemsep=0pt,parsep=0pt]

\item \textit{Satisfaction}: a read event has at least one corresponding
communication in $\rfcommunicationrelation$:
\bgroup\abovedisplayskip0.75\abovedisplayskip\belowdisplayskip0.5\belowdisplayskip\begin{eqntabular}[fl]{@{}l}
\forall i\in\interval[open left, open right]{0}{1+\lengthoftrace{\tau_p}}\suchthat\forall r\suchthat
({\eventsoftrace{\tau_p}}_i=r)\implies{}(\exists w\suchthat \catrfevent{w}{r}\in\rfcommunicationrelation)\ .
\defcondition(1.2em)<satisfaction>{satisfaction}{\execution}
\end{eqntabular}\egroup

\item \textit{Singleness}: a read event in the trace $\tau_p$  must have at
most one corresponding communication in $\rfcommunicationrelation$:
\bgroup\abovedisplayskip0.5\abovedisplayskip\belowdisplayskip0.5\belowdisplayskip\begin{eqntabular}[fl]{@{}l}
\forall r \ w \ w'\suchthat(\catrfevent{w}{r}\in\rfcommunicationrelation\wedge\catrfevent{w'}{r}\in\rfcommunicationrelation)\implies (w=w')\ .
\defcondition(1.3em)<singleness>{singleness}{\execution}
\end{eqntabular}\egroup
Note however that a read instruction can be repeated in a program loop and may
give rise to several executions of this instruction, each recorded by a unique
read event.\llstrut

\item \textit{Match}:  if a read reads from a write, then the variables read
and written must be the same:
\bgroup\abovedisplayskip0.5\abovedisplayskip\belowdisplayskip0.5\belowdisplayskip\begin{eqntabular}[fl]{@{}l}
(\catrfevent{\writeevent{\markertuple{p}{\ell}{\avar{x} \aassign \areg{r}}{\theta}}{v}}{\readevent{\markertuple{p'}{\ell'}{\areg{r}' \aassign \avar{x}'}{\theta'}}{\svar{x}{\theta}'}}\in\rfcommunicationrelation)
\nonumber\\
\qquad{}\implies{}(\avar{x}=\avar{x}')\ .
\defcondition<match>{match}{\execution}
\end{eqntabular}\egroup

\item \textit{Inception}: no communication is possible without the occurrence
of both the read and (maybe initial) write it involves:
\bgroup\abovedisplayskip0.5\abovedisplayskip\belowdisplayskip0.5\belowdisplayskip\begin{eqntabular}[fl]{@{\hspace*{-0.75em}}l}
\forall r \ w{\suchthat}{}
(\catrfevent{w}{r}\in\rfcommunicationrelation)\implies{}(\exists p\in\setofallprocessidentifiers,q\in\setofallprocessidentifiersstart{\suchthat}{} \defcondition(0.5em)<inception>{inception}{\execution}\\
\qquad \exists j\in\interval[open right]{0}{1+\lengthoftrace{\tau_p}},k\in\interval[open right]{0}{1+\lengthoftrace{\tau_q}}\suchthat {\eventsoftrace{\tau_p}}_j=r\wedge {\eventsoftrace{\tau_q}}_k=w)\ .
\nonumber
\end{eqntabular}\egroup
Note that this does not prevent a read to read from a future write.

\end{itemize}

\paragraph{Language-dependent conditions for \LISA $\!\!\!\!\!\!$} are as
follows:

\begin{itemize}[align=left,leftmargin=*,topsep=0pt,itemsep=0.pt,parsep=0pt]
\item \textit{Start}: the initial state of a trace $\trace_{p}$ should
be of the form:
\bgroup\abovedisplayskip0.5\abovedisplayskip\belowdisplayskip0.5\belowdisplayskip\begin{eqntabular}[fl]{@{\marge}l}
{\statesoftrace{\tau_p}}_{0}=\stateofprocess{\grlabel_p^0}{\infstamp{p}}{\LAMBDA{\areg{R}}0}{\emptyset}
\defcondition<zeroing>{zeroing}{\execution}
\end{eqntabular}\egroup
where $\grlabel_p^0$ is the entry label of process $p$ and $\infstamp{p}$ is a minimal stamp of $p$.

\item \emph{Next state}: if at point $k$ of a trace $\tau_p$ of process $p$ of
an execution $\execution
={\parallelexecution{\trace_{\startprocess}}{p\in\setofallprocessidentifiers}{}{\trace_p}{}{\rfcommunicationrelation}}$
the computation is in state
${\statesoftrace{\tau_p}}_{k-1}=\stateofprocess{\ell}{\theta}{\aenvironment}{\avaluation}$
then:
\begin{itemize}[align=left,leftmargin=*,topsep=0pt,itemsep=0.pt,parsep=0pt]
\item the next event must be generated by the instruction \defnonterminal{instr} $\triangleq$ $\attinstr{\texttt{P}}_p{\ell}$
at label $\ell$ of process $p$
\item the next event has the form ${\eventsoftrace{\tau_p}}_{k}$ =
$\mathfrak{e}\triple{{\markertuple{p}{\ell}{\nonterminal{instr}}{\theta}}}{\svar{x}{\theta}}{v}$
\item the next state
${\statesoftrace{\tau_p}}_{k}=\stateofprocess{\acontrol'}{\theta'}{\aenvironment'}{\avaluation'}$
has $\acontrol'=\ell'$ which is the label after the instruction
\nonterminal{instr}
\item the stamp $\theta'=\succstamp{p}(\theta)$ is larger, and 
\item the value $v$ as well as the new environment ${\aenvironment'}$ and
valuation ${\avaluation'}$ are computed as a function of the previous
environment ${\aenvironment}$, the valuation ${\avaluation}$, and the execution
$\execution$.  \end{itemize}
\end{itemize}

\noindent Formally: $\forall k\in\interval[open left,
open right]{0}{1+\lengthoftrace{\tau_p}}\suchthat\forall\acontrol' \ \aenvironment \ \aenvironment' \ \avaluation \ \avaluation' \ \theta \ \theta' \suchthat{}$
\bgroup\abovedisplayskip0.5\abovedisplayskip\belowdisplayskip0.5\belowdisplayskip\begin{eqntabular}[fl]{@{}l} 
({\statesoftrace{\tau_p}}_{k-1}=\stateofprocess{\ell}{\theta}{\aenvironment}{\avaluation}\wedge{}
{\eventsoftrace{\tau_p}}_{k}=\mathfrak{e}\triple{{\markertuple{p}{\ell}{\nonterminal{instr}}{\theta}}}{\svar{x}{\theta}}{v}
\wedge{}\nonumber\\
{\statesoftrace{\tau_p}}_{k}=\stateofprocess{\acontrol'}{\theta'}{\aenvironment'}{\avaluation'})
\implies{}(\acontrol'={\ell'}\wedge\theta'=\succstamp{p}(\theta)\wedge{}
\ifInvariance\nonumber\\\fi
\renumber{$v=\mathbold{v}(\aenvironment) \wedge{}\aenvironment'=\mathbold{\aenvironment}(v,\aenvironment)\wedge\mathbold{\avaluation}(v,\aenvironment,\avaluation,\execution,\avaluation'))\
.$}
\end{eqntabular}\egroup
\noindent We give the form of the next event ${\eventsoftrace{\tau_p}}_{k}$ for
each \LISA instruction:

\begin{itemize}[align=left,leftmargin=*,topsep=0pt,itemsep=0pt,parsep=0pt]
\item \textit{Fence} (\nonterminal{instr} = $\ell: {\finstra{\tagsequence}\
\bigl[\{\grlabel_1^0 \ldots \grlabel_1^m\}\ \{\grlabel_2^0 \ldots
\grlabel_2^q\}\bigr]}; \ell': \ldots$): 
\bgroup\abovedisplayskip0.5\abovedisplayskip\belowdisplayskip0.5\belowdisplayskip\begin{eqntabular}[fl]{@{}l}
{\eventsoftrace{\tau_p}}_{k}=\markerevent{\markertuple{p}{\ell}{\finstra{\tagsequence}\ \bigl[\{\grlabel_1^0 \ldots \grlabel_1^m\}\ \{\grlabel_2^0 \ldots \grlabel_2^q\}\bigr]}{\theta}}
\nonumber\\
(\aenvironment'=\aenvironment\wedge\avaluation'=\avaluation)\ .
\defcondition<fence>{fence}{\execution}
\end{eqntabular}\egroup

\item \textit{Register instruction} (\nonterminal{instr} =
$\ell: {\mova{{\areg{R}_1}}{\nonterminal{operation}}}; \ell': \ldots$):
\bgroup\abovedisplayskip0.5\abovedisplayskip\belowdisplayskip0.5\belowdisplayskip\begin{eqntabular}[fl]{@{}l}
{\eventsoftrace{\tau_p}}_{k}=\assignmentevent{\markertuple{p}{\ell}{\mova{{\areg{R}_1}}{\nonterminal{operation}}}{\theta}}{v}\defcondition<assignment>{assignment}{\execution}\\
(v=\evalexprenv[{\nonterminal{operation}}](\aenvironment,\avaluation)\wedge\aenvironment'=\environmentassignment{\aenvironment}{\areg{R}_1}{v}\wedge{}\avaluation'=\avaluation)\
.\nonumber
\end{eqntabular}\egroup
 where $\defevalexprenv[e](\aenvironment,\avaluation)$ is the evaluation of the
expression $e$ in the environment $\aenvironment$  and valuation $\avaluation$.

\item \textit{Write} (\nonterminal{instr} =
$\ell: {\winstra[\tagsequence]{\avar{x}}{}{\nonterminal{r-value}}}; \ell': \ldots$):
\bgroup\abovedisplayskip0.5\abovedisplayskip\belowdisplayskip0.5\belowdisplayskip
\begin{eqntabular}[fl]{@{}l}
{\eventsoftrace{\tau_p}}_{k}=\writeevent{\markertuple{p}{\ell}{\winstra[\tagsequence]{\avar{x}}{}{\nonterminal{r-value}}}{\theta}}{v}\defcondition<awrite>{awrite}{\execution}\\
(v=\evalexprenv[{\nonterminal{r-value}}](\aenvironment,\avaluation)\wedge{}\aenvironment=\aenvironment'\wedge{}\avaluation'=\avaluation)\ .\nonumber
\end{eqntabular}\egroup

\item \textit{Read} (\nonterminal{instr} =
$\ell: {\rinstra[\tagsequence]{{\areg{R}_1}}{\avar{x}}{}}; \ell': \ldots$):
\bgroup\abovedisplayskip0.5\abovedisplayskip\belowdisplayskip0.5\belowdisplayskip\begin{eqntabular}[fl]{@{}l}
{\eventsoftrace{\tau_p}}_{k}=\readevent{\markertuple{p}{\ell}{\rinstra[\tagsequence]{{\areg{R}_1}}{\avar{x}}{}}{\theta}}{\svar{x}{\theta}}\defcondition<aread>{aread}{\execution}\\
(\aenvironment'=\environmentassignment{\aenvironment}{{\areg{R}_1}}{\svar{x}{\theta}}{}\wedge{}
\exists{q\in\setofallprocessidentifiersstart}\suchthat\exists! j\in\interval[open right]{1}{1+\lengthoftrace{\tau_q}}\suchthat{}\nonumber\\
\quad\exists \ell'', \theta'',v\suchthat({\eventsoftrace{\tau_q}}_{j}={\writeevent{\markertuple{q}{\ell''}{\winstra[\tagsequence]{\avar{x}}{}{\nonterminal{r-value}}}{\theta''}}{v}}\wedge{}\nonumber\\
\qquad\catrfevent{{\eventsoftrace{\tau_q}}_{j}}{{\eventsoftrace{\tau_p}}_{k}}\in\rfcommunicationrelation\wedge{}
\avaluation'=\environmentassignment{\avaluation}{\svar{x}{\theta}}{v}))\ .\nonumber
\end{eqntabular}\egroup

\item \textit{RMW} (\nonterminal{instr} =
\rmw[\tagsequence]{\areg{r}}{\nonterminal{reg-instrs}}{\avar{x}}{}{}): for the begin  (\nonterminal{instr} =
${\grbeginrmw[\tagsequence]{x}}$) and end event (\nonterminal{instr} = ${\grendrmw[\tagsequence]{x}}$):
\bgroup\abovedisplayskip0.5\abovedisplayskip\belowdisplayskip0.5\belowdisplayskip\begin{eqntabular}[fl]{@{}l}
{\eventsoftrace{\tau_p}}_{k}=\markerevent{\markertuple{p}{\ell}{\nonterminal{instr}}{\theta}}{}\defcondition<rmw>{rmw}{\execution}\\
(
\aenvironment'=\aenvironment\wedge{}
\avaluation'=\avaluation)\ .\nonumber
\end{eqntabular}\egroup

\item \textit{Test} (\nonterminal{instr} =
$\ell: {\binstra[\tagsequence]{\nonterminal{operation}}{\grlabel_{\mskip-2mu
t}}{}}; \ell': \ldots$):
  \begin{itemize}[align=left,leftmargin=*,topsep=0pt,itemsep=0.pt,parsep=0pt]
  \item on the true branch:
\bgroup\abovedisplayskip0.5\abovedisplayskip\belowdisplayskip0.5\belowdisplayskip\begin{eqntabular}{l}
{\eventsoftrace{\tau_p}}_{k}=\testevent{\markertuple{p}{\ell}{\binstra[\tagsequence]{\nonterminal{operation}}{\grlabel_{\mskip-2mu t}}{}}{\theta}}{}\defcondition<test>{test}[t]{\execution}\\
(\satisfiable{\evalexprenv[\nonterminal{operation}](\aenvironment,\avaluation)\neq0}\wedge{}\acontrol'={}\grlabel_{\mskip-2mu t}{}\wedge{}\aenvironment'=\aenvironment\wedge\avaluation'=\avaluation)\nonumber
\end{eqntabular}\egroup

  \item on the false branch:
\bgroup\abovedisplayskip0.5\abovedisplayskip\belowdisplayskip0.5\belowdisplayskip\begin{eqntabular}{l}
{\eventsoftrace{\tau_p}}_{k}=\testevent{\markertuple{p}{\ell}{ \binstra[\tagsequence]{\mneg\nonterminal{operation}}{\grlabel_{\mskip-2mu t}}}{\theta}}{}\renumber{$\refcondition{test}[f]{\execution}$}\\
(\satisfiable{\evalexprenv[\nonterminal{operation}](\aenvironment,\avaluation)=0}\wedge{}\acontrol'=\ell'{}\wedge{}
\aenvironment'=\aenvironment\wedge\avaluation'=\avaluation)
\nonumber
\end{eqntabular}\egroup
  \end{itemize}
\end{itemize}

\subsubsection{Anarchic semantics} 
The anarchic semantics of a program
$\texttt{P}$ is
\bgroup\abovedisplayskip-1pt\belowdisplayskip-7pt
\begin{eqntabular*}{rcl}
\semantics<anarchictag>[\texttt{P}]&\triangleq&
\{\execution\in\setofallexecutions\mid \refcondition{start}{\execution} \wedge\ldots\wedge
\refcondition{test}{\execution}\}\ .
\end{eqntabular*}\egroup
\ifInvariance
\begin{theorem}
In an execution $\computation \times\rfcommunicationrelation\in\semantics<anarchictag>[\texttt{P}]$, the communication $\rfcommunicationrelation$ uniquely determines the computation $\computation$.
\end{theorem}
\fi

\subsubsection{\cat specification of a weakly consistent semantics} \label{sec:cat-specification-weakly-consistent-semantics}
The candidate execution abstraction
$\defalphacandidateexecution(\execution)$ abstracts the execution
$\execution=\computation \times\rfcommunicationrelation$ into a candidate execution $\alphacandidateexecution(\execution)$ = $\cattuple{\mathit{e}}{\mathit{po}}{\mathit{rf}}{\mathit{iw}}{\mathit{fw}}$ where ${\mathit{e}}$ is the set of events (partitionned into fence, read, write, \ldots\ events), ${\mathit{po}}$ is the program order (transitively relating successive events on a trace of each process), ${\mathit{rf}}$ = $\rfcommunicationrelation$ is the set of communications, and ${\mathit{iw}}$ is the set of initial write events. Then we define
$\defalphacandidateexecution(\semantics)\triangleq\{\pair{\execution}{\alphacandidateexecution(\execution)}\mid
\execution\in\semantics\}$ and
$\defalphacat[\Hcom](\setofcandidateexecutions)\triangleq\{\execution,\communicationrelation\mid
\pair{\execution}{\candidateexecution}\in\setofcandidateexecutions\wedge\pair{\texttt{allowed}}{\communicationrelation}\in\catsemantics[\Hcom](\candidateexecution)\}$
where the consistence $\defcatsemantics[\Hcom](\candidateexecution)$ of a
candidate execution $\candidateexecution$ for a \cat consistency model $\Hcom$ is defined in \cite{AlglaveCousotMaranget-cat-HSA-2015} and returns communication relations $\communicationrelation$ psecifying communication constraints on communication events. The
analytic semantics of a program $\texttt{P}$ for a \cat specification $\Hcom$ is then $\semantics[\texttt{P}]$ $\triangleq$
$\alphacat[\Hcom]\comp\alphacandidateexecution(\semantics<anarchictag>[\texttt{P}])$.

\endgroup

\begingroup
\section{An overview of \LISA}

\newcommand{\Peterson}{\hyperlink{Peterson}{Peterson}\xspace}

\subsection{Example}

To illustrate \LISA we use \Peterson's algorithm, given
in Figure~\ref{fig:Peterson-algorithm}.

The algorithm uses three shared variables \texttt{F1}, \texttt{F2} and \texttt{T}:
\begin{itemize}[align=left,leftmargin=*,topsep=0.25\topsep,itemsep=0.25\itemsep]
\item two shared flags, \texttt{F1} for the first process \texttt{P0} (resp.
\texttt{F2} for the second process \texttt{P1}), indicating that the process
\texttt{P0} (resp. \texttt{P1}) wants to enter its critical section, and
\item a turn \texttt{T} to grant priority to the other process: when \texttt{T}
is set to \texttt{1} (resp. \texttt{2}), the priority is given to \texttt{P0}
(resp. \texttt{P1}).
\end{itemize}

Let's look at the process \texttt{P0}: \texttt{P0} busy-waits before entering
its critical section (see the \texttt{do} instruction at line \texttt{3:})
until (see the \texttt{while} clause at line \texttt{6:}) the process
\texttt{P1} does not want to enter its critical section (\viz{}, when {\tt
F2=false}, which in turn means $\neg$\texttt{R1=true} thanks to the read at
line \texttt{4:}) or if \texttt{P1} has given priority to \texttt{P0} by
setting turn \texttt{T} to \texttt{1}, which in turn means that \texttt{R2=1}
thanks to the read at line \texttt{5:}.

\noindent\begin{figure}[t]\noindent\begin{minipage}[t]{0.49\textwidth}\bgroup\ttfamily\footnotesize\abovedisplayskip0pt\belowdisplayskip0pt\begin{eqntabular*}{@{\hskip-3mm}l@{}}
\begin{tabular}[t]{@{\hskip-3mm}l@{}l@{}}
0:&\{~F1 = false;~F2 = false;~T = 0; \}
\end{tabular}\\[-0.5ex]\noindent\begin{tabular}[t]{@{\hskip-3mm}l@{}l@{$\;$}||@{$\;$}l@{}l}
\rlap{P0:}&                       & P1: & \\
1: &F1 = true                     & 10: &F2 = true;\\
2: &T = 2                         & 11: &T = 1;\\
3: &do                            & 12: &do  \\
4: &~~~R1 = F2                    & 13: &~~~R3 = F1;\\
5: &~~~R2 = T                     & 14: &~~~R4 = T;\\
6: &while R1 $\wedge$ R2 $\neq$ 1 & 15: &while R3 $\wedge$ R4 $\neq$ 2;  \\
7: &skip (* CS1 *)                & 16: &skip (* CS2 *)\\
8: &F1 = false                    & 17: &F2 = false;\\
9: &                              & 18: &
\end{tabular}
\end{eqntabular*}
\egroup
\end{minipage}\begin{minipage}[t]{0.5\textwidth}
\footnotesize
\begin{verbatim}
LISA Peterson
{ }
P0                      | P1                     ;
L1:  w[] F1 1           | L10: w[] F2 1          ;
L2:  w[] T 2            | L11: w[] T 1           ;
L4:     r[] r1 F2       | L13:    r[] r3 F1      ;
L5:     r[] r2 T        | L14:    r[] r4 T       ;
L20: mov r7 (neq r2 1)  | L31: mov r6 (neq r4 2) ;
L21: mov r9 (and r1 r7) | L32: mov r8 (and r3 r6);
L6:  b[] r9 L4          | L15: b[] r8 L13        ; 
L8:  w[] F1 0           | L17: w[] F2 0          ;
L9:                     | L18:                   ;
\end{verbatim}
\end{minipage}
\caption{\protect\hypertarget{Peterson}{Peterson} algorithm---in \protect\LISA\label{fig:Peterson-algorithm}}
\end{figure}

\paragraph{LISA code}
Let's read it together; our algorithm is composed of:
\begin{itemize}[align=left,leftmargin=*,topsep=0.25\topsep,itemsep=0.25\itemsep]
\item a prelude at line {\tt 0:}, between curly brackets, which
initialises the variables {\tt F1} and {\tt F2} to {\tt false} and the variable
{\tt T} to {\tt 0}. By default initialisation is to {\tt 0} ({\tt false});
\item two processes, each depicted as a column; let's detail the first process,
on the left-hand side: at line {\tt L1:} we write {\tt 1} ({\tt true}) to the shared variable
{\tt F1}---the \LISA syntax for writes is ``{\tt w[] x e}'' where {\tt x} is a
variable and {\tt e} an expression over registers, whose value is written to
{\tt x}. At line {\tt L2} of we write {\tt 2} to {\tt T}. At line {\tt 3:} of \hyperlink{Peterson}{Peterson} algorithm, we see a
{\tt do} instruction which ensures that we iterate the instructions at
lines {\tt 4} and {\tt 5} until the condition expressed at line {\tt 6}
(\viz{}, {\tt R1 }$\wedge${ \tt R2 $\neq$ 1}) is false.  In the \LISA translation,
at line {\tt L4:} we read the variable {\tt F2} and write its value into register {\tt R1}, and at line {\tt L5:} we read the variable {\tt T} and write its value into register {\tt R2}. At lines {\tt L20:} and {\tt L21:} we locally compute
the value {\tt r1} $\wedge${ \tt r2 $\neq$ 1}) in local register  {\tt r9}. At line
{\tt L6:} the branch instruction {\tt b[] r9 L4} branches to {\tt L4:} if {\tt r9} is 1 ({\tt true}) \ie the loop body is iterated once more and continue in sequence when {\tt r9} is 0 ({\tt false}) \ie the loop body is exited and the critical section is entered. At line  {\tt L8:} we write {\tt 0} ({\tt false}) to {\tt F1}.
\end{itemize}\endgroup

\begingroup
\makeatletter
\newcommand{\setofallpythiavariablesoperator}[1]{\mathfrak{P}\def\@setofallpythiavariables{#1}\ifx\@setofallpythiavariables\@empty\else(#1)\fi}
\newcommand{\defsetofallpythiavariables}[1]{\hypertarget{hyper:setofallpythiavariables}{\setofallpythiavariablesoperator{#1}}}
\newcommand{\setofallpythiavariables}[1]{\hyperlink{hyper:setofallpythiavariables}{\setofallpythiavariablesoperator{#1}}}
\makeatother

\vskip-1mm

\ifInvariance
We present here a little language called \LISA (Litmus Instruction Set
Architecture). Its vocation is purely pedagogical at the moment, with an
ambition to be quite minimal. It is supported by the \textsf{herd7}
tool \cite{herd7}. To illustrate this section we will again use \Peterson's algorithm, given
in Figure~\ref{fig:Peterson-algorithm}.
\fi

\subsection{Syntax}

\paragraph{\LISA programs $\!\!\!\!\!\!$}
$\texttt{P}=\{\texttt{P}_{\startprocess}\}\aprogram{\texttt{P}_0}{\ldots}{\texttt{P}_{n-1}}$
on shared variables $\avar{x}\in\defvarof[P]$ contain:
\begin{itemize}[align=left,leftmargin=*,topsep=0pt,itemsep=0pt,parsep=0pt]
\item a prelude $\texttt{P}_{\defstartprocess}$ assigning initial values to shared
variables. In the case of Peterson algorithm in Figure~\ref{fig:Peterson-algorithm}, the
prelude at line {\tt 0:} assigns the value {\tt false} to both variables {\tt
F1} and {\tt F2}, and the value $0$ to {\tt T}. This initialization to {\tt 0} ({\tt false}) is implicit in the \LISA translation;
  \item processes $\texttt{P}_{0}$ \dots $\texttt{P}_{n-1}$ in parallel; each
process:
\begin{itemize}[align=left,leftmargin=*,topsep=0pt,itemsep=0pt,parsep=0pt]
    \item has an identifier
$p\in\setofallprocessidentifiers[\texttt{P}]\triangleq\interval[open
right]{0}{n}$; in the case of Peterson we have  used
\texttt{P0} for the first process (on the left) and \texttt{P1} for the second
process (on the right);
    \item has local registers (\eg{} $\areg{R0, R1}$); registers are assumed to
be different from one process to the next; if not we make them different by
affixing the process identifier like so: \texttt{($p$:$\areg{R}$)});
    \item is a sequence of instructions.  
\end{itemize}
\end{itemize}

\paragraph{Instructions $\!\!\!\!\!$} can be:
\begin{itemize}[align=left,leftmargin=*,topsep=0pt,itemsep=0pt,parsep=0pt]
\item \emph{register instructions}
\mova{{\areg{R1}}}{\nonterminal{operation}}, where the \defnonterminal{operation}
has the shape \oparith[\nonterminal{op}]{{\areg{R2}}}{\nonterminal{r-value}}:
  \begin{itemize}[align=left,leftmargin=*,topsep=0pt,itemsep=0.pt,parsep=0pt]
  \item the operator \defnonterminal{op} is arithmetic (\eg{} \oparith[add]{}{}, \oparith[sub]{}{}, \oparith[mult]{}{}) or boolean (\eg{} \opbool[eq]{}{}, \opbool[neq]{}{}, \opbool[gt]{}{}, \opbool[ge]{}{}); 
  \item ${\areg{R1}}$ and $\areg{R2}$ are local registers;
  \item \defnonterminal{r-value} is either a local register or a constant;
  \end{itemize}

\item \emph{read instructions} \rinstra[\tagsequence]{{\areg{R}}}{\avar{x}}{}
initiate the reading of the value of the shared variable $\avar{x}$ and write it into the local
register ${\areg{R}}$;

\item \emph{write instructions}
\winstra[\tagsequence]{\avar{x}}{}{e} initiate the writing of the
value of the register expression $e$ into the shared variable $\avar{x}$;

\item \emph{branch instructions}
\binstra[\tagsequence]{\nonterminal{operation}}{\grlabel_{\mskip-2mu t}} branch
to label ${\grlabel_{\mskip-2mu t}}$ if the ${\nonterminal{operation}}$ has
value \texttt{true} and go on in sequence otherwise;

\item \emph{fence instructions} $\finstra{\tagsequence}\ \bigl[\{\grlabel_1^0
\ldots \grlabel_1^m\}\ \{\grlabel_2^0 \ldots \grlabel_2^q\}\bigm]$. The
optional sets of labels $\{\grlabel_1^0 \ldots \grlabel_1^m\}$ and
$\{\grlabel_2^0 \ldots \grlabel_2^q\}$ indicate that the fence $\finstra{}$
only applies between instructions in the first set and instructions in the
second set.  The semantics of the fence $\finstra{}$ as
applied to the instructions at $\grlabel_1^0$, \ldots, or $\grlabel_1^m$ and
$\grlabel_2^0$, \ldots, $\grlabel_2^q$, is to be defined in a \cat
specification; 

\item \begin{minipage}[t]{0.65\columnwidth}\emph{read-modify-write instructions} (RMW)
$\rmw[\tagsequence]{\areg{R}}{\nonterminal{reg-instrs}}{\avar{x}}{}{}$ are 
translated into a sequence of instructions delimited by markers
$\grbeginrmw[\tagsequence]{x}$ and $\grendrmw[\tagsequence]{x}$ as
shown opposite. $\defnonterminal{reg-instrs}$ is a sequence of register instructions, which
computes in $\areg{R}$ the new value assigned to the shared variable
$\avar{x}$. 
\end{minipage}\hfil
\begin{minipage}[t]{0.30\columnwidth}\abovedisplayskip0pt\belowdisplayskip0pt\begin{eqntabular*}[fl]{@{\quad}l}
\grbeginrmw[\tagsequence]{x}\semicolonsymbol\\
\rinstra[\tagsequence]{{\areg{R}}}{\avar{x}}{}\semicolonsymbol\\
\nonterminal{reg-instrs}\semicolonsymbol\\
\winstra[\tagsequence]{\avar{x}}{}{{\areg{R}}}\semicolonsymbol\\
\grendrmw[\tagsequence]{x}
\end{eqntabular*} 
\end{minipage}
\ulstrut Any semantics requirement on RMWs, such as the fact that there can
be no intervening write to \avar{x} between the read and the write of the RMW,
has to be ensured by a \cat specification.
\end{itemize}

Instructions can be labelled (\ie{} be preceded by a control label $\ell$) to
be referred to in branches or fences for example. Labels are unique; if not we
make them different by affixing the process identifier like so: $p$:$\grlabel$.
$\attinstr{\texttt{P}}_p{\ell}$ is the instruction at label $\ell\in\setofallalabels{p}$ of process $p$ of program \texttt{P}.
Moreover, instructions can bear tags $\tagsequence$ (to model for example C++
release and acquire annotations).  
\ifInvariance\else
Scopes are special tags (\eg to model \eg Nvidia PTX~\cite{abd15} and HSA~\cite{HSA-Foundation-PSAS-2015}), whose semantics must be defined in a \cat specification. Scopes can be organized in the scope tree. The events created by a process will be automatically tagged by the scope of that process. For example the \LISA program \texttt{MP-scoped} in Figure~\ref{fig:mp-scoped} is augmented by a scope tree.
\begin{figure}[!h]
\begin{quote}
\IfFileExists{mp-scoped.litmus}{
\verbatiminput{mp-scoped.litmus} 
}{
\verbatiminput{../cat-semantics/mp-scoped.litmus} 
}
\end{quote}
\vspace*{-3mm}
\caption{The example \texttt{MP-scoped}} \label{fig:mp-scoped}
\end{figure}
The scope tree \texttt{scopes: (system (wi P0) (wi P1))} in
Figure~\ref{fig:mp-scoped} specifies that the threads~\texttt{P0}
and~\texttt{P1} reside in two different \emph{scope instances} of
level~\texttt{wi}.  By contrast, still as specified by the scope tree, there is
one scope instance of level~\texttt{system} and both threads reside in this
common instance.
\fi

\endgroup

\begingroup
\section{The anarchic true parallel formal semantics with separated communications of \protect\LISA}\label{sec:anarchic-semantics-LISA}
\let\section\subsection
\let\subsection\subsubsection
\let\subsubsection\paragraph
We now instantiate the general definition of an anarchic semantics of a parallel program of Section \ref{sec:Anarchic-semantics} to the case of the \LISA language.

We introduce the anarchic true parallel semantics $\semantics<anarchictag>$ with separated unconstrained communications in Section \ref{sec:true-parallel-symbolic-semantics} and provide ground value and symbolic instances of the anarchic semantics for the little language \LISA, in Section \ref{appendix:LISA}. The abstraction of the executions to candidate executions is specified in Section \ref{sec:abstraction-to-candidate-execution}. This is used in Section \ref{sec:abstraction-to-semantics-with-WCM} to specify the semantics of a program with a \cat weak consistency model $M$ constraining communications. The analytic semantics is the anarchic semantics with separated  communications $\semantics<anarchictag>$ constrained by  a \cat\ the weak consistency model. It is analytic in that it separates the definition of the computational
semantics $\semantics<anarchictag>$ from the communication semantics specified by a \cat\ specification. 

The definition of the anarchic semantics is in two parts. The first part in Section \ref{sec:true-parallel-symbolic-semantics} is language independent.
The second part in Section \ref{appendix:LISA} is language dependent for \LISA.

\section{The anarchic true parallel symbolic and ground valued semantics}\label{sec:true-parallel-symbolic-semantics}
The anarchic true parallel semantics avoids interleaving thanks to a concurrent representation of execution traces of processes with separate communications.
The anarchic semantics can be ground when taking values in a ground set $\setofallvaluesground$ ($\setofallintegers$ for \LISA).
The anarchic semantics can also be
symbolic when values are symbolic expressions in the symbolic variables denoting communicated values (called pythia variables). This generalises symbolic execution \cite{DBLP:journals/cacm/King76} to
finite and infinite executions of parallel programs with weak consistency
models. Since in a true parallel semantics there is no notion of
global time, there is also no notion of instantaneous value of shared variables. The only knowledge on the value of shared
variables is local, when a process has read a shared variable. We use a pythia variable to denote the value which is read (at some local time
symbolically denoted by a stamp). The usage of
pythia variable is not strictly necessary in the ground semantics. It is useful  in the symbolic semantics to denote values
symbolically. It is indispensable in invariance proof methods to give names to values as required in formal logics.

\subsection{Semantics}
The semantics $\semantics[\texttt{P}]$ of a concurrent program $\texttt{P}=\aprogram{\texttt{P}_0}{\ldots}{\texttt{P}_{n-1}}\in\setofallaprograms$ 
with $n\geqslant 1$ processes ${\texttt{P}_0},{\ldots},{\texttt{P}_p},{\ldots},{\texttt{P}_{n-1}}$ 
identified by their pids $p\in\setofallprocessidentifiers[\texttt{P}]\triangleq\interval[open right]{0}{n}$ 
is a set of executions $\semantics[\texttt{P}]\in\wp(\setofallexecutions[\texttt{P}])$. We omit $\sqb{\texttt{P}}$
when it is understood from the context.

\subsection{Computations and executions}\label{sec:Executions}

A computation $\computation\in\defsetofallcomputations\triangleq\setofalltraces(\startprocess)\times\prod_{p\in\setofallprocessidentifiers}\setofalltraces(p)\times\setofalltraces(\finishprocess)$ has the form $\computation$ = $\parallelexecution{\trace_{\startprocess}}{p\in\setofallprocessidentifiers}{}{\trace_p}{\trace_{\finishprocess}}{}$
where $\startprocess\not\in\setofallprocessidentifiersfinish$ is the initialization/prelude process, $\finishprocess\not\in\setofallprocessidentifiersstart$ is the finalization/postlude process, $\trace_{\startprocess}\in\setofalltraces(\startprocess)$ is an execution trace of the initialization process, $\trace_p\in\setofalltraces(p)$ are execution traces of the processes $p\in\setofallprocessidentifiers$, and $\trace_{\finishprocess}\in\setofalltraces(\finishprocess)$ is an execution trace of the finalization process.

An execution $\execution\in\defsetofallexecutions\triangleq\setofallcomputations\times\wp(\setofallcommunicationevents)$ has the form $\computation\times\rfcommunicationrelation$ = $\parallelexecution{\trace_{\startprocess}}{p\in\setofallprocessidentifiers}{}{\trace_p}{\trace_{\finishprocess}}{\rfcommunicationrelation}$  where $\rfcommunicationrelation\in\wp(\setofallcommunicationevents)$ is the communication relation (read-from). $\setofallcommunicationevents$ will be defined in Sect.\nbsp\ref{sec:Communications} as the set of read-write event pairs on the same shared variable with the same value.
 
\subsection{Traces}\label{sec:Traces}

A finite non-empty trace $\trace_p\in\setofalltraces<finite>$, $p\in\setofallprocessidentifiersstartfinish$
is a finite sequence of computation steps $\pair{\epsilon}{\sigma}$, 
represented as $\eventstate{\epsilon}{\sigma}$ (with event $\epsilon\in\setofallevents(p)$ and next state $\sigma\in\setofallstates(p)$), of the form
\begin{eqntabular*}{rcl}
\trace_p & = &\eventstate{\epsilon_1}{\sigma_1}\eventstate{\epsilon_2}{\sigma_2}\eventstate{\epsilon_3}{\sigma_3}\ldots\sigma_{n-1}\eventstate{\epsilon_{m}}{\sigma_{m}}\\
     & = &\langle\eventstate{\epsilon_{k}}{\sigma_{k}}\mid k\in\interval[open right]{1}{1+m}\rangle
\end{eqntabular*}
such that $\epsilon_1=\startevent$ is the start event. A computation step $\eventstate{\epsilon}{\sigma}$ represents the atomic execution of an action that (1) creates event $\epsilon$ and (2) changes the state to $\sigma$.
The trace $\trace_p$ has length $\lengthoftrace{\trace_p}=m$. For brevity, we  use the more traditional form $\trace_p = {\sigma_1}\eventstate{\epsilon_2}{\sigma_2}\eventstate{\epsilon_3}{\sigma_3}\ldots\sigma_{m-1}\eventstate{\epsilon_{m}}{\sigma_{m}}$
since the first event $\epsilon_1$ is always the start event $\epsilon_1=\startevent$. We say that event $\epsilon_k$, state $\sigma_{k}$, and step $\eventstate{\epsilon_{k}}{\sigma_{k}}$ are ``at point $k\in\interval[open right]{1}{1+\lengthoftrace{\trace_p}}$ of $\trace_p$''.

An infinite trace $\trace_p\in\setofalltraces<infinite>$, $p\in\setofallprocessidentifiers$ has the form 
\begin{eqntabular*}{rcl}
\trace_p & = & \eventstate{\epsilon_1}{\sigma_1}\eventstate{\epsilon_2}{\sigma_2}\eventstate{\epsilon_3}{\sigma_3}\ldots\sigma_{n-1}\eventstate{\epsilon_{n}}{\sigma_{n}}\ldots\\
     & = &\langle\eventstate{\epsilon_{k}}{\sigma_{k}}\mid k\in\setofallstrictlypositive\rangle
\end{eqntabular*}
and has length $\lengthoftrace{\tau}=\infty$ such that $1+\infty=\infty$\footnote{$\,$With this convention, the domain of a finite or infinite trace is $\interval[open right]{1}{1+\lengthoftrace{\tau}}$ (where $\infty$ is the first infinite ordinal). If $\lengthoftrace{\tau}$ is finite then $\interval[open right]{1}{1+\lengthoftrace{\tau}}$ = $\interval{1}{\lengthoftrace{\tau}}$. Else $\lengthoftrace{\tau}$ is infinite so $1+\lengthoftrace{\tau}=1+\infty=\infty$ in which case $\interval[open right]{1}{1+\lengthoftrace{\tau}}$
= $\interval[open right]{1}{\infty}$.}. It is traditionally written $\tau = {\sigma_1}\eventstate{\epsilon_2}{\sigma_2}\eventstate{\epsilon_3}{\sigma_3}\ldots\sigma_{n-1}\eventstate{\epsilon_{n}}{\sigma_{n}}\ldots$ where $\epsilon_1=\startevent$.

We let $\setofalltraces\colsep{\triangleq} \setofalltraces<finite>\cup \setofalltraces<infinite>$ be the set of (finite or infinite) traces.

We define the sequence of states of a trace $\tau$ as 
\begin{eqntabular*}{rcl}
\statesoftrace{\tau}&\triangleq&\sigma_1\sigma_2\sigma_3\ldots\sigma_{n-1}\sigma_{n}[\ldots]
\end{eqntabular*}
and its sequence of events of a trace $\tau$ as  
\begin{eqntabular*}{rcl}
\eventsoftrace{\tau}\colsep{\triangleq}\epsilon_1\epsilon_2\epsilon_3\ldots\epsilon_{n-1}\epsilon_{n}[\ldots].
\end{eqntabular*}
(We write $[\ldots]$ when the elements $\ldots$ are optional \eg to have a single notation for both finite and infinite traces.) So a trace has the form $\tau=\langle\eventstate{\eventsoftrace{\tau}_{k}}{\statesoftrace{\tau}_{k}}\mid k\in\interval[open right]{1}{1+\lengthoftrace{\tau}}\rangle$. 

\subsection{Stamps}\label{sec:stamps}
Stamps are used to ensure that events in a trace of a process are unique.
\begin{itemize}[itemsep=3pt]

\item $\theta\in\defsetofallstamps{p}$ the stamps (or postmarks) of process $p\in\setofallprocessidentifiersstartfinish$ uniquely identify events when executing process $\texttt{P}_p$. Different processes have different stamps so for all $p,q\in\setofallprocessidentifiersstartfinish$ if $p\neq q$ then $\setofallstamps{p}\cap\setofallstamps{q}=\emptyset$;

\item $\displaystyle\theta\in\setofallstamps{}\colsep{\triangleq}\bigcup_{p\in\setofallprocessidentifiersstartfinish}\setofallstamps{p}$.

\end{itemize}
We assume that stamps are totally ordered per process.
\begin{itemize}[itemsep=3pt]

\item $\mathord{\defleqstamp{p}}\in\wp(\setofallstamps{p}\times\setofallstamps{p})$ totally orders $\setofallstamps{p}$ ($\defltstamp{p}$ is the strict version);

\item $\defsuccstamp{p}\in\setofallstamps{p}\rightarrow\setofallstamps{p}$ is the successor function (\issuchthat $\theta\ltstamp{p}\succstamp{p}(\theta)$, not necessarily the immediate successor);

\item $\definfstamp{p}\in\setofallstamps{p}$ is the smallest stamp for process $p$ (\issuchthat $\forall\theta\in\setofallstamps{p}\suchthat\infstamp{p}\ltstamp{p}\theta$).
\end{itemize}
These hypotheses allow us to generate a sequence of unique stamps when building traces. Starting from any stamp (\eg $\infstamp{p}$), a trace where events are successors by $\succstamp{p}$ is guaranteed to have unique stamps (since moreover different processes have different stamps). Stamps are otherwise unspecified and can be defined freely. For example we can use process labels and/or loop counters.

\subsection{Equivalence of executions and semantics up to stamp renaming}
We consider executions $\setofallexecutions\quotient{\traceequivalence}$ up to the isomorphic renaming $\traceequivalence$ of stamps. The renaming is an equivalence relation.
\begin{eqntabular}[fl]{@{\marge}l}
\parallelexecution{\trace_{\startprocess}}{p\in\setofallprocessidentifiers}{}{\trace_p}{\trace_{\finishprocess}}{\rfcommunicationrelation}
\traceequivalence
\parallelexecution{\trace'_{\startprocess}}{p\in\setofallprocessidentifiers}{}{\trace'_p}{\trace'_{\finishprocess}}{\rfcommunicationrelation'}\colsep{\triangleq}\nonumber\\
\renumber{$\exists\varrho\in\setofallstamps{}\twoheadleftrightarrow\setofallstamps{}\suchthat
\quad\forall p\in\setofallprocessidentifiersstartfinish\suchthat\tau'_p=\varrho(\tau_p)
\wedge\rfcommunicationrelation'=\varrho(\rfcommunicationrelation)$\ .}
\end{eqntabular}
where $\twoheadleftrightarrow$ denotes an isomorphism and $\varrho$ is homomorphically extended from stamps to events, states, traces, and communications containing these stamps.

We also consider semantics $\semantics\quotient{\traceequivalence}$ up to the isomorphic renaming $\traceequivalence$ of stamps by defining semantics to be equivalence when identical up to isomorphic stamp renaming of their traces.
\begin{eqntabular*}[fl]{@{\marge}rcl}
\semantics\traceequivalence\semantics{\mskip0.75mu}'&\triangleq&\bigl\{\execution\quotient{\traceequivalence}\bigm|\execution\in\semantics\bigr\}=\bigl\{\execution'\quotient{\traceequivalence}\bigm|\execution'\in\semantics{\mskip0.75mu}'\bigr\}\ .
\end{eqntabular*}

\subsection{Shared variables, registers, denotations, and data}\label{sec:Data}
\begin{itemize}[itemsep=3pt]

\item $\avar{x}\in\defvarof[\texttt{P}]\subseteq\defsetofallavariables$: shared variables of program $\texttt{P}$;

\item $\areg{r}\in\defsetofallaregisters{p}$:  local registers of process $p\in\setofallprocessidentifiersfinish$; 

\item $\acst{d}\in\defsetofalladenotations$:  set of all data/value denotations;

\item $\acst{0}\in\setofalladenotations$: we assume that registers and, in absence of prelude, shared variables are implicitly initialized by a distinguished initialisation value denoted $\acst{0}$;

\item We let $\setofallvalues$ be the set of all ground or symbolic values manipulated by programs and $\adenotedvalue{}\in\setofalladenotations\rightarrow\setofallvalues$ is the interpretation of data/value denotations into 
data/value.

\end{itemize}

\subsection{Pythia variables}\label{sec:Media-variables}
\begin{itemize}[itemsep=3pt]

\item Pythia variables $\svar{x}{\theta}$ are used to ``store/record'' the value of a shared variable $\avar{x}$ when accessing this variable\footnote{similarly to single assignment languages but not SSA which is a static abstraction.}. Pythia variables $\svar{x}{\theta}$ can be thought of as addresses, locations, L-values in buffers, or channels, communication lines to indirectly designate the R-value of a shared variable observed by a specific \texttt{read} event stamped $\theta$. The R-value designated by an L-value $\svar{x}{\theta}$ may be unknown. For example, the actual value may be assigned in the future \eg in the thin-air case\footnote{Again this situation may have to be rejected by a \ifInvariance $\Hcom/M$ \else \cat\ \fi specification.}.
\begin{eqntabular}[fl]{@{\marge}rcl}
\defsetofallsymbolicvariables(p) &\triangleq&\{ \svar{x}{\theta} \mid \avar{x}\in\setofallavariables\wedge \theta\in\setofallstamps{p}\}\renumber{\begin{tabular}[t]{l}pythia variables of process $p\in\setofallprocessidentifiersfinish$\\at stamp $\theta$\end{tabular}}
\\
\setofallsymbolicvariables &\triangleq&\bigcup_{p\in\setofallprocessidentifiers}\setofallsymbolicvariables(p) \renumber{pythia variables.}
\end{eqntabular}
Note that we designate pythia variables $\svar{x}{\theta}$ via a stamp $\theta$ to guarantee their uniqueness in the trace.
Observe that we use three different kinds of variables. Program variables like local registers $\areg{r}$ and shared variables $\avar{x}$ which appear in the syntax and semantics, meta or mathematical variables like $p$ or $\theta$ which are used in the definitions, theorems, and proofs but are not part of the syntax and semantics (only the objects denoted by these mathematical variables do appear in the syntax and semantics), and pythia variables like \eg $\svar{x}{0}$, $\svar{y}{2}$, \etc which appear in the semantics. So $\svar{x}{\theta}$ is one of these pythia variables where the mathematical variable $\theta$ denotes any of the stamps in $\setofallstamps{p}$ \eg $0$, $1$, \etc. 

\end{itemize}

\subsection{Expressions}\label{sec:Expressions}\label{sec:Evaluation-expressions-in-environments}

In the symbolic semantics, symbolic values will be expressions on pythia variables $\setofallsymbolicvariables$ (and possibly other symbolic variables),
in which case $\setofallvalues=\setofallaexpressions[\setofallsymbolicvariables]$ and the interpretation $\adenotedvalue{}$ is the identity.
In the ground semantics, values belong to a ground set $\setofallvaluesground$ ($\setofallintegers$ for \LISA).

\begin{itemize}[itemsep=3pt]

\item $e,e_1,e_2,\ldots\in\defsetofallaexpressions[V]$: set of all mathematical expressions over variables $v\in V$, $e\gis\acst{d}\gor v\gor e_1 \mathbin{\obar} e_2$ (where $\obar$ is a mathematical (\eg arithmetical) operator);

\item $b,b_1,b_2,\ldots\in\defsetofallbexpressions[V]$:  set of all boolean expressions over variables $v\in V$, $b\gis e_1 \olessthan e_2\gor b_1 \varodot b_2 \gor \oneg b$ (where $\olessthan$ is a comparison, $\varodot$ is a boolean operator, and $\oneg$ is negation).

\item The evaluation $\defevalexprenv[e](\aenvironment,\avaluation)$ of a mathematical expression $e\in\setofallaexpressions[V]$ over variables $v\in V$ in an environment
$\aenvironment\in V\rightarrow\setofallvalues\cup\setofallsymbolicvariables$ mapping variables $v\in V$ to their value $\aenvironment(v)\in\setofallvalues$ or to a pythia variable
and a valuation $\avaluation\in \setofallsymbolicvariables\rightarrow\setofallvalues$ mapping pythia variables ${\svar{x}{\theta}}\in\setofallsymbolicvariables$ to their value $\avaluation({\svar{x}{\theta}})\in\setofallvalues$
is defined by structural induction on $e$ as
\begin{eqntabular}{@{\marge}rcl@{\quad}L}
\evalexprenv[\acst{d}](\aenvironment,\avaluation)&\triangleq&\adenotedvalue{\acst{d}}&\defdefinition{evalexprenv}
\\
\evalexprenv[v](\aenvironment,\avaluation)&\triangleq&\aenvironment(v)&symbolic semantics or ground value 
\nonumber\\
&&&semantics when $\aenvironment(v)\in\setofallvalues$ is ground
\nonumber\\
&\triangleq&\evalexprenv[\aenvironment(v)](\aenvironment,\avaluation)&ground value semantics when $\aenvironment(v)\in\setofallsymbolicvariables$
\nonumber\\
&&&is not ground
\nonumber\\
\evalexprenv[{\svar{x}{\theta}}](\aenvironment,\avaluation)&\triangleq&\avaluation({\svar{x}{\theta}})&
\nonumber\\
\evalexprenv[e_1 \obar e_2](\aenvironment,\avaluation)
&\triangleq&
\evalexprenv[e_1](\aenvironment,\avaluation)
\mathbin{\adenotedvalue{\obar}}
\evalexprenv[e_2](\aenvironment,\avaluation)
\nonumber
\end{eqntabular}
where $\adenotedvalue{\acst{d}}$ in the interpretation of the constant $\acst{d}$ and ${\adenotedvalue{\obar}}$ is the interpretation of the mathematical operator ${\obar}$.

\item  The evaluation $\defevalbexprenv[b](\aenvironment,\avaluation)\in\setofallbooleans\colsep{\triangleq}\{\true,\false\}$ of a boolean expression $b\in\setofallbexpressions[V]$ in environment
$\aenvironment\in V\rightarrow\setofallvalues$ is defined as
\begin{eqntabular}{@{\marge}rcl@{\quad}L}
\evalbexprenv[e_1 \olessthan e_2](\aenvironment,\avaluation)&\triangleq&\evalexprenv[e_1](\aenvironment,\avaluation)
\mathrel{\adenotedvalue{\olessthan}}
\evalexprenv[e_2](\aenvironment,\avaluation)
\defdefinition{evalbexprenv}\\
\evalbexprenv[b_1 \varodot b_2](\aenvironment,\avaluation)&\triangleq&
\evalbexprenv[b_1](\aenvironment,\avaluation)
\mathbin{\adenotedvalue{\varodot}}
\evalbexprenv[b_2](\aenvironment,\avaluation)
\nonumber\\
\evalbexprenv[\oneg b](\aenvironment,\avaluation)&\triangleq&
\mathbin{\adenotedvalue{\oneg}}
\evalbexprenv[b](\aenvironment,\avaluation)\ .
\nonumber
\end{eqntabular}
where ${\adenotedvalue{\olessthan}}$ is the interpretation of the comparison operator $\olessthan$, ${\adenotedvalue{\varodot}}$ is the interpretation of the boolean operator $\varodot$, and ${\adenotedvalue{\oneg}}$ is the interpretation of the negation $\oneg$.

\end{itemize}

\subsection{Events}\label{sec:Events}
\subsubsection{\bfseries Start event.}\quad

\begin{itemize}[itemsep=3pt]

\item $\defstartevent$ is the start event at the beginning of traces.
\begin{eqntabular*}[fl]{@{\marge}rcl@{\quad}L}
\setofallstartevents&\triangleq&\{\startevent\}.
\end{eqntabular*}
The $\startevent$ event is not indispensable. It is used to represent uniformly traces as sequences of computation steps \ie pairs of an event
and a state. Otherwise a trace would be a sequence of states separated by events. This conventional representation is dissymmetric which is why we choose to have a $\startevent$ event. 

\end{itemize}

\subsubsection{\bfseries Computation events.}\quad
\begin{itemize}[itemsep=3pt]
\newcommand{\myboxlength}{224.91385pt}

\item $c(p)\in\defsetofallcomputationevents(p)$ is the set of computation events of process $p\in\setofallprocessidentifiers$. In the case of \LISA, the computation events $\setofallevents\sqb{\nonterminal{proc}}$ of a process $\nonterminal{proc}$ are defined in Fig.\nbsp\ref{fig:lisa-proc}. They can be marker events (defined for fences in Fig.\nbsp\ref{fig:lisa-fences} and for read-modify-write instructions in Fig~\ref{fig:lisa-rmws}), 
register assignment events (defined in Fig.\nbsp\ref{fig:lisa-regs}), and test events (defined in Fig.\nbsp\ref{fig:lisa-branches}).

\end{itemize}

\subsubsection{\bfseries Read events.}\quad
\begin{itemize}[itemsep=3pt]
\newcommand{\myboxlength}{210.71222pt}

\item $r(p,\avar{x},v)\in\setofallreadevents(p,\avar{x},v)$ is the set of read events of process $p\in\setofallprocessidentifiers$ reading value $v\in\setofallvalues$ from shared variable $\avar{x}\in\varof[\textup{\texttt{P}}]$;

\item $r(\finishprocess,\avar{x},v)\in\setofallreadevents(\finishprocess,\avar{x},v)$ is the set of final read events reading value $v\in\setofallvalues$ from  shared variable $\avar{x}\in\varof[\textup{\texttt{P}}]$;
\renewcommand{\myboxlength}{157.70604pt}

\item $\displaystyle r(p)\in\defsetofallreadevents(p)\colsep{\triangleq}\bigcup_{\avar{x}\in\varof[\textup{\texttt{P}}]}\bigcup_{v\in\setofallvalues}\setofallreadevents(p,\avar{x},v)\lLstrut$ is the set of read events of process $p\in\setofallprocessidentifiersfinish$;

\end{itemize}
Read events for \LISA are defined in Fig.\nbsp\ref{fig:lisa-reads}. Each read event stamped $\theta$ uses a pythia variable $\svar{x}{\theta}$  to store the value read/to be read by the read event. The pythia variable $\svar{x}{\theta}$ is always unique in a trace $\tau$ since the stamp $\theta$ of the event is assumed (by 
the forthcoming condition \refcondition{uniqueness}{\tau}) to be unique on that trace $\tau$.

\subsubsection{\bfseries Write events.}\quad
\begin{itemize}[itemsep=3pt]

\item $w(p,\avar{x},v)\in\setofallwriteevents(p,\avar{x},v)$  is the set of write events of process $p\in\setofallprocessidentifiersstart$ writing value $v\in\setofallvalues$ into shared variable $\avar{x}\in\varof[\textup{\texttt{P}}]$;
\newcommand{\myboxlength}{209.38951pt}

\item $w(\startprocess,\avar{x},v)\in\setofallwriteevents(\startprocess,\avar{x},v)$  is the set of initial write events of value $v\in\setofallvalues$ into shared variable $\avar{x}\in\varof[\textup{\texttt{P}}]$;

\item $\displaystyle w(p)\in\defsetofallwriteevents(p)\colsep{\triangleq}\bigcup_{\avar{x}\in\setofallavariables}\bigcup_{v\in\setofallvalues}\setofallwriteevents(p,\avar{x},v)$ is the set of write events of process $\texttt{P}_p$, $p\in\setofallprocessidentifiersstart$;

\end{itemize}
For the \LISA language, write events are defined in Fig.\nbsp\ref{fig:lisa-writes}.

\subsubsection{\bfseries Events.}\quad
\begin{itemize}[itemsep=3pt]
\item {$\epsilon, \epsilon(p)\in\defsetofallevents(p)\colsep{\triangleq}\setofallcomputationevents(p)\cup\setofallreadevents(p)\cup\setofallwriteevents(p)$} is the of computation events of process $p\in\setofallprocessidentifiersstartfinish$;

\item {$\displaystyle\setofallevents[$p$]\colsep{\triangleq}\bigcup_{p\in\setofallprocessidentifiersstartfinish}\setofallevents(p)$} is the set of events of program $\texttt{P}=\aprogram{\texttt{P}_1}{\ldots}{\texttt{P}_n}\in\setofallaprograms$; For the \LISA programs, it is defined in Fig.\nbsp\ref{fig:lisa-progs}.

\end{itemize}

\subsection{States}\label{sec:States}
\ifInvariance Invariance proofs involve sets of states. These states\else States \fi are the control state and the memory state mapping variables to their value. We use an environment to record the value of local registers, as is classical. However, the instantaneous values of shared variables are unknown to processes. We use instead pythia variables to record the values read by processes. The valuation maps these pythia variables to the value of the corresponding shared variable at the time when it was read (\ie which in general is not the instantaneous value, otherwise unknown). 

\ifInvariance
We would like to evaluate the $\Hcom/M$ WCM specification using the invariant only (without concretizing to all possible executions satisfying the invariant). The invariant provides the communications for that purpose.
This also requires to order events per process according to the program order. Traditionally invariants provide no order information on states (although this information can be incorporated indirectly in states via auxiliary variables like loop counters). 
\else
We would like all events to be distinct, ordered per process, but not interprocesses.
\fi
Therefore we had stamps to states (knowing by the hypotheses of Section \ref{sec:stamps} that there are ordered per process according to the program order of that process).

\begin{itemize}[itemsep=3pt]

\item The states of the $\startprocess$ and $\finishprocess$ processes are meaning less and so are $\bullet$ where $\setofallstates(\startprocess)=\setofallstates(\finishprocess)=\{\bullet\}$.

\item The states $\sigma\in\setofallstates(p)$ of process $p\in\setofallprocessidentifiers$ have the form $\sigma$ = $\defstateofprocess{\acontrol}{\theta}{\aenvironment}{\avaluation}$ where

\begin{itemize}[itemsep=3pt]
\item $\acontrol\in\setofallalabels{p}$ is the current control label/program point of process $p$ (we have $\attdone[\texttt{P}](p)\acontrol$ which is $\true$ if and only if $\acontrol$ is the last label of process $p$ which is reached if and only if process $p$ does terminate); 

\item $\theta\in\setofallstamps{p}$ is the stamp of the state in process $p$;

\item $\aenvironment\in\setofallaenvironments{p}\triangleq\setofallaregisters{p}\rightarrow\setofallvalues\cup\setofallsymbolicvariables/\setofallaexpressions[\setofallsymbolicvariables(p)]$ is an environment mapping local registers $\areg{r}\in\setofallaregisters{p}$ of process $p$ to their ground/symbolic value. This is a map since the process registers are known statically;

\item $\avaluation\in\defsetofallavaluations{p}\triangleq\setofallvalues/\setofallsymbolicvariables(p)\prightarrow\setofallaexpressions[\setofallsymbolicvariables(p)]$ is a valuation mapping pythia variables of process $p$ to their ground/symbolic value. This is a partial map since the pythia variables are known dynamically. We write $\domain(\avaluation)$ for the domain of definition of $\avaluation$ (initially $\emptyset$ at execution start).

\end{itemize}
\end{itemize}

\subsection{Well-formed traces}

The execution $\execution=\parallelexecution{\trace_{\startprocess}}{p\in\setofallprocessidentifiers}{}{\trace_p}{\trace_{\finishprocess}}{\rfcommunicationrelation}\in\setofallexecutions$ is well-formed under the following well-formedness conditions. The finalisation $\trace_{\finishprocess}$ 
\ifInvariance
has been omitted in the paper for brevity, but 
\fi
it is necessary to specify the outcome on the program computations (in conjunction with a \cat specification of how the final communications should be performed).

\setcounter{equation}{1}
\begin{itemize}[itemsep=3pt]

\item \textit{Start}: traces $\tau\in\{\tau_p\mid p\in\setofallprocessidentifiers\}\cup\{\trace_{\startprocess},{\trace_{\finishprocess}}\}\in\wp(\setofalltraces)$ must all start with a unique start event on the trace.
\begin{eqntabular}{@{}l@{}}
\eventsoftrace{\tau}_{1}=\startevent\wedge\forall i\in\interval[open left, open right]{1}{1+\lengthoftrace{\tau}}\suchthat\eventsoftrace{\tau}_{i}\neq\startevent\ .
\defcondition<start>{start}{\execution}\end{eqntabular}

\item \textit{Uniqueness}: the stamps of events in a trace $\tau\in\{\tau_p\mid p\in\setofallprocessidentifiers\}\cup\{\trace_{\startprocess},{\trace_{\finishprocess}}\}\in\wp(\setofalltraces)$ must be unique on the trace (the initial write events as well as the
final read and communication events are unique per shared variable and so do not need stamps to be distinguished).
\begin{eqntabular}{l@{\qquad}}
\forall i,j\in\interval[open right]{1}{1+\lengthoftrace{\tau}}\suchthat
(i\neq j\wedge \eventsoftrace{\tau_i},\eventsoftrace{\tau_j}\not\in(\setofallreadevents(\finishprocess)\cup\setofallwriteevents(\startprocess)))
\defcondition<uniqueness>{uniqueness}{\execution}\\
\qquad{}\implies{}\stampof(\eventsoftrace{\tau_i})\neq\stampof(\eventsoftrace{\tau_j})\ .
\nonumber
\end{eqntabular}
(It immediately follows that all events occurring on a trace are unique. Moreover, the symbolic variable $\svar{x}{\theta}$ in any read event $\tau_k=\readevent{\markertuple{p}{\ell}{\areg{r} \aassign \avar{x}}{\theta}}{\svar{x}{\theta}}$ of a trace $\tau$ is unique on that trace.)

\item \textit{Initialisation}: all shared variables $\avar{x}\in\setofallavariables$ are assumed to be initialized once and only once to a value $v_{\avar{x}}\in\setofallvalues$ in the sequential prelude (or to $v_{\avar{x}}=\adenotedvalue{\acst{0}}$ by default).

\begin{eqntabular}[fl]{@{\marge}l@{}l}
\forall\avar{x}\in\setofallavariables\suchthat&
(\exists \tau_1,\tau_2\suchthat {\trace_{\startprocess}} = \tau_1 \:(\writeevent{\markertuple{\startprocess}{\startlabel}{\avar{x} \aassign e}{\theta}}{v_{\avar{x}}})\: \tau_2)
\wedge{}\defcondition<initialisation>{initialisation}{\execution}
\\
&(\forall \tau_1,\tau_2,\tau_3\suchthat ({\trace_{\startprocess}} = \tau_1 \:\epsilon\: \tau_2 \: \writeevent{\markertuple{\startprocess}{\ell}{\avar{x} \aassign e}{\theta}}{v_{\avar{x}}}\: \tau_3)\nonumber\\
&\quad{}\implies{}(\epsilon\in\setofallwriteevents(\startprocess,\avar{y})\wedge \avar{y}\neq\avar{x}))\ .
\nonumber\end{eqntabular}

\item \textit{Finalisation}: all shared variables $\avar{x}\in\setofallavariables$ are finally read once and only once in the postlude and their final value is stored in a fresh register $\areg{r}_{\avar{x}}$.
\begin{eqntabular}[fl]{@{\marge}l}
(\forall\avar{x}\in\setofallavariables\suchthat
(\exists \tau_1,\tau_2\suchthat{\trace_{\finishprocess}}= \tau_1 \:\readevent{\markertuple{\finishevent}{\finishlabel}{\areg{r}_{\avar{x}} \aassign \avar{x}}{\theta}}{\svar{x}{\theta}}\: \tau_2)
\wedge{}
\renumber{\refcondition{initialisation}[f]{\execution}}
\\
(\forall\tau_1,\tau_2,\tau_3\suchthat (\tau = \tau_1 \:\readevent{\markertuple{\finishevent}{\ell}{\areg{r}_{\avar{x}} \aassign \avar{x}}{\theta}}{\svar{x}{\theta}}\: \tau_2 \: \epsilon'\: \tau_3)){}\implies{}\nonumber\\
\quad((\epsilon'\in\setofallreadevents(\finishprocess,\avar{y})\wedge\avar{y}\neq\avar{x})))\ .\nonumber
\end{eqntabular}
\global\def\finalisation{{\normalfont finalisation}\xspace}

\item \textit{Maximality}: finite trace $\tau_p$ of a process
$p\in\setofallprocessidentifiers$ must be maximal \ie must describe a process
which execution is finished. Note that infinite traces are maximal by
definition, hence we do not need to include them in the present maximality
condition.   
 
\begin{eqntabular}[fl]{@{\marge}rcl@{}}
\forall
p\in\setofallprocessidentifiers\suchthat(\tau_p\in{\setofalltraces}^+)&\implies&(
\exists \ell\in\setofallalabels{p}\suchthat \exists
\theta\in\setofallstamps{p}\suchthat \exists
\aenvironment\in\setofallaenvironments{p}\suchthat \exists
\avaluation\in\setofallavaluations{p}\suchthat
\defcondition{maximality}{\execution} \\
&&\qquad\qquad{\statesoftrace{\tau_p}}_{\lengthoftrace{\tau}}=\stateofprocess{\ell}{\theta}{\aenvironment}{\avaluation}
\wedge \attdone[\texttt{P}](p)\ell)\ .\nonumber
\end{eqntabular} (\ie the control state of the last state of the trace is at
the end of the process, as indicated by $\attdone[\texttt{P}](p)\ell$, which is
language dependent, and defined for \LISA in Sect.\nbsp\ref{appendix:LISA}.)
\end{itemize}

\subsection{Well-formed communications}

\subsubsection{\bfseries Communications.}\label{sec:Communications}\quad
\begin{itemize}[itemsep=3pt]
\newcommand{\myboxlength}{246.37553pt}

\item$c(p,w)\in\defsetofallcommunicationevents(p,w)$ is the set of communications of process $p\in\setofallprocessidentifiersfinish$ reading from the write $w$.
\begin{eqntabular}[fl]{@{\marge}rcl@{\quad}L}
c(p,w(q,\avar{x},v))&\gis&\catrfevent{w(q,\avar{x},v)}{r(p,\avar{x},v')} & \renumber{communication}
\end{eqntabular}
read event $r(p,\avar{x},v')$ of process $p\in\setofallprocessidentifiers$ or final read ($p=\finishprocess$) reads the value $v'$ of $\avar{x}$ from write event $w(q,\avar{x},v)$ of the same or another process $q\in\setofallprocessidentifiers$ or from initial write ($q$ = $\startprocess$) where $\satisfiable{v=v'}$.

\item $c(p)\in\setofallcommunicationevents(p)\colsep{\triangleq}\bigcup_{w\in\setofallwriteevents}\setofallcommunicationevents(p,w)$ is the set of communications for process $p\in\setofallprocessidentifiersfinish$;

\item $c(w)\in\setofallcommunicationevents(w)\colsep{\triangleq}\bigcup_{p\in\setofallprocessidentifiersfinish}\setofallcommunicationevents(p,w)$ is the set of communications satisfied by write event $w$;

\item $c\in\setofallcommunicationevents[\texttt{P}]\colsep{\triangleq}\bigcup_{p\in\setofallprocessidentifiersfinish}\setofallcommunicationevents(p)$ is the set of communications of program $\texttt{P}=\aprogram{\texttt{P}_1}{\ldots}{\texttt{P}_n}\in\setofallaprograms$;

\item {$\setofallcommunicationevents\colsep{\triangleq}\bigcup_{\texttt{P}\in\setofallaprograms}\setofallcommunicationevents[\texttt{P}]$} is the set of all communications.

\end{itemize}
Notice that communications are not stamped (precisely because we do not want to
impose any notion of time between communications). If a stamp is needed to uniquely identify a communication, we can
use the one of the read event involved in the communication since it is unique
by \refcondition{satisfaction}{\execution} and \refcondition{singleness}{\execution}.

\subsubsection{\bfseries Well-formed communications.}\label{sec:Well-formed-communications}

\quad To be well-formed, an execution $\execution=\parallelexecution{\trace_{\startprocess}}{p\in\setofallprocessidentifiers}{}{\trace_p}{\trace_{\finishprocess}}{\rfcommunicationrelation}\in\setofallexecutions$ must have communications $\rfcommunicationrelation$ satisfying the following conditions.

\begin{itemize}[itemsep=3pt]

\item \textit{Satisfaction}: a read event in the trace
$\tau_p\in\setofalltraces$ of a process $p\in\setofallprocessidentifiersfinish$
must have at least one corresponding communication in $\rfcommunicationrelation$
(since writes are fair \ie become ultimately readable and there is always an
initial readable write to initialize variables). We impose this condition on
reads to avoid the case where a read never reads anything, which would block the
execution.
\begin{eqntabular}[fl]{@{\marge}l}
\forall i\in\interval[open left, open right]{1}{1+\lengthoftrace{\tau_p}}\suchthat\forall r\in\setofallreadevents(p)\suchthat
({\eventsoftrace{\tau_p}}_i=r)\implies(\exists w\in\setofallwriteevents\suchthat \catrfevent{w}{r}\in\rfcommunicationrelation)\ .
\defcondition<satisfaction>{satisfaction}{\execution}
\end{eqntabular}

\item \textit{Singleness}: a read event in the trace $\tau_p\in\setofalltraces$ of a process $p\in\setofallprocessidentifiersfinish$  must have at most one corresponding communication in $\rfcommunicationrelation$.
\begin{eqntabular}[fl]{@{\marge}l}
\forall r\in\setofallreadevents(p)
\suchthat\forall w,w'\in\setofallwriteevents\suchthat(\catrfevent{w}{r}\in\rfcommunicationrelation\wedge\catrfevent{w'}{r}\in\rfcommunicationrelation)\implies
(w=w')\ .
\defcondition<singleness>{singleness}{\execution}
\end{eqntabular}
(Note however that a single read action (\ie atomic instruction in \LISA),  can be repeated in a program loop and may give rise to
several executions of this action, each recorded by unique read events (each one reading from one, possibly different, past or future write event.)

\item \textit{Match}:  if a read event in the trace $\tau\in\setofalltraces$ reads from a write event, then the variables read and written must be the same.
\begin{eqntabular}[fl]{@{\marge}l}
(\catrfevent{\writeevent{\markertuple{p}{\ell}{\avar{x} \aassign \areg{r}}{\theta}}{v}}{\readevent{\markertuple{p'}{\ell'}{\areg{r}' \aassign \avar{x}'}{\theta'}}{\svar{x}{\theta}'}}\in\rfcommunicationrelation)
{}\implies
(\avar{x}=\avar{x}')\ .
\defcondition<match>{match}{\execution}
\end{eqntabular}

\item \textit{Inception}: no communication is possible without the occurrence
of both the read and write events it involves (the write may be an initial
one). 
\begin{eqntabular}[fl]{@{\marge}l}
\forall r\in\setofallreadevents(p)\suchthat\forall w\in\setofallwriteevents\suchthat{}
\defcondition<inception>{inception}{\execution}\\
\quad(\catrfevent{w}{r}\in\rfcommunicationrelation)\implies{}(\exists p\in\setofallprocessidentifiersfinish,q\in\setofallprocessidentifiersstart\suchthat{} \nonumber\\
\qquad \exists j\in\interval[open right]{1}{1+\lengthoftrace{\tau_p}},k\in\interval[open right]{1}{1+\lengthoftrace{\tau_q}}\suchthat {\eventsoftrace{\tau_p}}_j=r\wedge {\eventsoftrace{\tau_q}}_k=w)\ .
\nonumber
\end{eqntabular}
(Note that this does not prevent a read to read from a future write.)

\end{itemize}

\subsection{Well-formed execution}
An execution $\execution=\parallelexecution{\trace_{\startprocess}}{p\in\setofallprocessidentifiers}{}{\trace_p}{\trace_{\finishprocess}}{\rfcommunicationrelation}\in\setofallexecutions$ is well-formed if and only if it satisfies all conditions \refcondition{start}{\execution} to \refcondition{inception}{\execution}. This leads to the definition of the semantic domain.
\begin{eqntabular}{rcl@{\qquad\qquad}}
\defsetofalltracesemantics&\triangleq&\{\semantics\in\wp({\setofalltraces\quotient{\traceequivalence}})\mid\forall\execution\in\semantics\suchthat\refcondition{start}{\execution}\wedge \ldots\wedge \refcondition{inception}{\execution}\}
\label{def:setofalltracesemantics}
\end{eqntabular}
Moreover, for a particular programming language, computation events on
process traces must be generated in program order. This well-formedness condition 
has to be specified for each programming language \eg \refcondition{zeroing}{\execution} to \refcondition{test}{\execution} below for \LISA. 

\subsection{Anarchic semantics}
Observe that if $\forall i\in\Delta.\semantics{}_i\in\setofalltracesemantics$ then $\bigcup_{i\in\Delta}\semantics{}_i\in\setofalltracesemantics$
and $\bigcap_{i\in\Delta}\semantics{}_i\in\setofalltracesemantics$ proving that $\sextuple{\setofalltracesemantics}{\subseteq}{\emptyset}{\semantics(\anarchictag)}{\cup}{\cap}$ is a complete sublattice of $\sextuple{\wp(\setofalltraces\quotient{\traceequivalence})}{\subseteq}{\emptyset}{\setofalltraces\quotient{\traceequivalence}}{\cup}{\cap}$. So $\setofalltracesemantics$ has an infimum $\emptyset$
and a supremum $\semantics(\anarchictag)\triangleq\Bigl(\bigcup_{\semantics\in\setofalltracesemantics}\semantics\Bigr)\in\setofalltracesemantics$ called the anarchic semantics. 

\section{Litmus Instruction Set Architecture (\LISA)}\label{appendix:LISA}

In this section we present a little language that we call \LISA, for
\emph{Litmus Instruction Set Architecture}. We provide the syntax, symbolic and
ground semantics of \LISA.

\subsection{Programs}
The semantics of programs is defined by an attribute grammar
\cite{DBLP:conf/waga/Knuth90} (see \cite{DBLP:journals/csur/Paakki95} for an introduction)
given in Figure~\ref{fig:lisa-progs}. 

The program may have a sequential prelude, \viz, a set of initial 
assignments of values $v_{\avar{x}}$ to shared variables $\avar{x}$, 
a case covered by \refcondition{initialisation}{\execution} and by 
$p=\startprocess$ in  Fig.\nbsp\ref{fig:lisa-writes}. We omit the 
specification of the syntax and semantics of the prelude.
In absence of prelude, for an empty prelude \texttt{\{\}}, or in absence of a 
specific initialization of some shared variables, the shared variables are assumed 
to be implicitly initialized to $\adenotedvalue{\acst{0}}$.

Local registers of processes are assumed to be implicitly initialized to
$\adenotedvalue{\acst{0}}$, as shown in Fig.\nbsp\ref{fig:lisa-proc}.

The \LISA version used by the \herdseven tool also offers the possibility
to initialize registers ${\areg{r}}$ in the prelude. 
The register is designated
by $\mathopen{\texttt{(}}p\colonsymbol{\areg{r}}\mathclose{\texttt{)}}$  where
$p\in\setofallprocessidentifiers$ is their process identifier.   This is
equivalent to moving these initializations at the beginning of the
corresponding process $p$ since they will be executed after the default
initialization of registers to $0$.

The \LISA version used by the \herdseven tool also offers the possibility to define a postlude to
check whether there exists a finite execution $\tau$ or for all finite
executions $\tau$ a condition on the trace $\tau$ does hold. This boolean
condition may involve the final value of registers
$\mathopen{\texttt{(}}p\colonsymbol{\areg{r}}\mathclose{\texttt{)}}$ evaluated
by $\evalexprenv[\areg{r}](\tau,\lengthoftrace{\tau})$ where $p$ is a process
identifier. This boolean condition may also involve the final value of shared
variables $\avar{x}$ as stored in register ${\areg{r}_{\avar{x}}}$ in
Fig.\nbsp\ref{fig:lisa-reads} where $p=\finishprocess$.

One can optionally specify a scope tree to be used in the \cat communication specification. 

The attributes of a program $\nonterminal{prog}$ are 
its number $\attpid(\nonterminal{prog})$ of processes, 
the events $\setofallevents\sqb{\nonterminal{prog}}$ that can be generated by the instructions of the program, 
and its semantics $\attsemantics{\nonterminal{prog}}$. 

For clarity, some attributes are left implicit such as the local labels $\setofallalabels{p}$, the local registers $\setofallaregisters{p}$
of process $p\in\interval[open right]{0}{\attpid(\nonterminal{prog})}$. These attribute definitions including restrictions such as uniqueness can be easily added to the attribute grammar.

The anarchic semantics $\semantics<anarchictag>[\nonterminal{body}]$ is the set of all executions satisfying conditions $\attwf{\nonterminal{body}}$ expressing that the execution $\execution$ must correspond to an execution of the $\nonterminal{body}$. Moreover all executions $\execution$ in this anarchic semantics
$\attsemantics{\nonterminal{prog}}$ must satisfy well-formedness conditions \refcondition{start}{\execution}, \ldots, \refcondition{inception}{\execution} as expected by the \cat semantics.

\begin{figure}[!ht]

\begin{pcframed}[Programs]
\begin{eqntabular*}[fl]{@{\quad}rcl}
\defnonterminal{prog}&\in&\setofnonterminal{proc}\label{eq:def:[a]program}\\
\nonterminal{prog}&{}\grdef{}&\nonterminal{body}{}\gror{}\texttt{\{\}}\ \nonterminal{body}\\
&{}\gror{}&\nonterminal{prelude}\ \nonterminal{body}\\
&{}\gror{}&\nonterminal{prelude}\ \nonterminal{body}\ \nonterminal{postlude}\\
&{}\gror{}&\nonterminal{prelude}\ \nonterminal{body}\ \texttt{scopes:}\:\nonterminal{scope-tree}\\
&{}\gror{}&\nonterminal{prelude}\ \nonterminal{body}\ \texttt{scopes:}\:\nonterminal{scope-tree}\ \nonterminal{postlude}\ 
\end{eqntabular*}
In all cases, $\llet\ n=\attpid(\nonterminal{body})\ \lin$ \hfill (last process identifier in $\nonterminal{body}$)
\begin{itemize}[itemsep=3pt]

\item $\attpid(\nonterminal{prog})\colsep{\triangleq}n+1$; \hfill (number of processes)

\item $\setofallprocessidentifiers[\nonterminal{body}]\colsep{\triangleq}\interval{0}{n}$; \hfill (program process identifiers)

\item $\attdone[\,\nonterminal{prog}]\colsep{\triangleq}\attdone[\nonterminal{body}]$; \hfill (last control label check)

\item ${\attinstr{\nonterminal{prog}}}\colsep{\triangleq}{\attinstr{\nonterminal{prelude}}}\cup
{\attinstr{\nonterminal{body}}}$\hfill (instructions)

\item $\setofallevents\sqb{\nonterminal{prog}}\colsep{\triangleq}\setofallevents\sqb{\nonterminal{body}}$; \hfill (program events)

\item $\displaystyle\attsemantics{\nonterminal{prog}}\colsep{\triangleq}
\{\parallelexecution{\trace_{\startprocess}}{p=1}{n}{\trace_p}{\trace_{\finishprocess}}{\rfcommunicationrelation}\mid\attwf{\nonterminal{body}}(\parallelexecution{\trace_{\startprocess}}{p=1}{n}{\trace_p}{\trace_{\finishprocess}}{\rfcommunicationrelation})
\wedge{}$\\
\mbox{}\hfill$\displaystyle\forall\textit{rf}\in\rfcommunicationrelation\suchthat
\neg\attwf{\nonterminal{body}}(\parallelexecution{\trace_{\startprocess}}{p=1}{n}{\trace_p}{\trace_{\finishprocess}}{\rfcommunicationrelation\setminus\{\textit{rf}\}})\}
$
\\
(\ie $\rfcommunicationrelation$ is minimal).

\item $\attsemantics{\nonterminal{prog}}\in\setofalltracesemantics$.

\end{itemize}
\end{pcframed}
\caption{\LISA programs \label{fig:lisa-progs}}
\end{figure}

\FloatBarrier

\subsection{Tags and scope trees} 

Instructions can bear tag sequences $\tagsequence$, that are given a
semantics within \cat specifications \cite{AlglaveCousotMaranget-cat-HSA-2015}.  
Scope trees can be used to describe program architecture-dependent features 
for the \cat specification \cite{AlglaveCousotMaranget-cat-HSA-2015}. These tag
sequences and scope tree  are not involved at all in the program 
anarchic semantics of \LISA. They are added in the form 
$\tagsequencemark{\tagsequence}$ to events forwarded to \cat. If there is no scope tree declaration, 
the trivial scope tree \texttt{(trivial P0 P1 \ldots P${}_{n-1}$)} is 
used where $n=\attpid(\nonterminal{body})$ is the number of processes in the program.
A process can appear at most once in a scope tree. Again scope trees are information for a 
\cat specification, see \cite{AlglaveCousotMaranget-cat-HSA-2015}.

\subsection{Parallel processes}
Parallel processes are given in Figure~\ref{fig:lisa-procs}. These process identifiers $\attpid$ are
defined to be $0$ to $n-1$ from left to right where $n=\attpid(\nonterminal{body})$ is the number
of processes in the parallel program. The program events $\setofallevents\sqb{\nonterminal{body}}$
and well-formedness conditions $\attwf{\nonterminal{body}}$ are collected.

\begin{figure}[!ht]
\begin{pcframed}[Parallel processes]
\begin{eqntabular*}[fl]{@{\quad}rcl@{\qquad}l}
\defnonterminal{body}&\in&\setofnonterminal{body}\\
\nonterminal{body}&\grdef&\phantom{{}\gror{}}\nonterminal{proc}\\
&&\qquad\attpid(\nonterminal{body})\colsep{\triangleq}\attpid(\nonterminal{proc})\colsep{\triangleq}0\\
&&\qquad\setofallprocessidentifiers[\nonterminal{proc}]\colsep{\triangleq}\setofallprocessidentifiers[\nonterminal{body}]\\
&&\qquad\attinstr{\nonterminal{body}}\colsep{\triangleq}\attinstr{\nonterminal{proc}}\\
&&\qquad\attdone[\,\nonterminal{body}](0)\colsep{\triangleq}\LAMBDA{\ell}(\ell=\attafter(\nonterminal{proc}))\\
&&\qquad\setofallevents\sqb{\nonterminal{body}}\colsep{\triangleq}\setofallevents\sqb{\nonterminal{proc}}\\
&&\qquad\attwf{\nonterminal{body}}\colsep{\triangleq}\LAMBDA{\execution}\attwf{\nonterminal{proc}}\execution\\
&&{}\gror{}\nonterminal{body}_1\parallel\nonterminal{proc}\\
&&\qquad\llet\ n=\attpid(\nonterminal{body}_1)\ \lin\\
&&\qquad\quad\attpid(\nonterminal{body})\colsep{\triangleq}\attpid(\nonterminal{proc})\colsep{\triangleq}n+1\\
&&\qquad\quad\setofallprocessidentifiers[\nonterminal{body}$_1$]\colsep{\triangleq}\setofallprocessidentifiers[\nonterminal{body}],\quad\setofallprocessidentifiers[\nonterminal{proc}]\colsep{\triangleq}\setofallprocessidentifiers[\nonterminal{body}]\\
&&\qquad\quad\attdone[\,\nonterminal{body}](p)\colsep{\triangleq}\LAMBDA{\ell}\si p\leqslant n\alors\attdone[\,\nonterminal{body}_1](p)\ell\sinon(\ell=\attafter(\nonterminal{proc}))\fsi\\
&&\qquad\quad\attinstr{\nonterminal{body}}\colsep{\triangleq}\attinstr{\nonterminal{body}_{1}}\cup\attinstr{\nonterminal{proc}}\\
&&\qquad\quad\setofallevents\sqb{\nonterminal{body}}\colsep{\triangleq}\setofallevents\sqb{\nonterminal{body}_1}\cup\setofallevents\sqb{\nonterminal{proc}}\\
&&\qquad\quad\attwf{\nonterminal{body}}\colsep{\triangleq}\LAMBDA{\execution}\attwf{\nonterminal{body}_1}{\execution}\wedge{}
\attwf{\nonterminal{proc}}{\execution}
\end{eqntabular*} 
\end{pcframed}

\caption{\LISA processes \label{fig:lisa-procs}}
\end{figure}

\FloatBarrier

\subsection{Processes} 
Processes are given in Figure~\ref{fig:lisa-proc}. Each process $\nonterminal{proc}$ of the program body $\nonterminal{body}$ has a unique
process identifier attribute $p=\attpid(\nonterminal{proc})$. Each instruction $\nonterminal{instr}$ of a process has a label $\attat(\nonterminal{instr})$
before and a label after $\attafter(\nonterminal{instr})$ that instruction. The label $\grlabel$ is 
after the last instruction of the list of instructions of the process. Each process with pid $p$ has
a unique entry label $\attat(\nonterminal{instrs})$ which is the one of its first instruction and is
where the process $p$ execution must start from, so $\attat(\nonterminal{instrs})$ is the label of the first control state.

\setcounter{equation}{9}\begin{figure}[!ht]
\begin{pcframed}[Processes]
\begin{eqntabular}[fl]{@{\quad}rcl}
\defnonterminal{proc}&\in&\setofnonterminal{proc}\nonumber\\
&\grdef&\nonterminal{instrs}\ \grlabel\colonsymbol\nonumber\\
&&\quad\llet\ p = \attpid(\nonterminal{proc})\ \land\nonumber\\
&&\qquad\attpid(\nonterminal{instrs})\colsep{\triangleq}p\nonumber\\
&&\qquad\setofallprocessidentifiers[\nonterminal{instrs}]\colsep{\triangleq}\setofallprocessidentifiers[\nonterminal{proc}]\nonumber\\
&&\qquad\attinstr{\nonterminal{proc}}\colsep{\triangleq}\attinstr{\nonterminal{instrs}}\nonumber\\
&&\qquad\setofallevents\sqb{\nonterminal{proc}}\colsep{\triangleq}\setofallevents\sqb{\nonterminal{instrs}}\nonumber\\
&&\qquad\attafter(\nonterminal{proc})\colsep{\triangleq}\attafter(\nonterminal{instrs})\colsep{\triangleq}\grlabel\nonumber\\
&&\qquad\attwf{\nonterminal{proc}}\colsep{\triangleq}
\LAMBDA{\execution}
\llet\ \execution = {\parallelexecution{\trace_{\startprocess}}{p\in\setofallprocessidentifiers}{}{\trace_p}{\trace_{\finishprocess}}{\rfcommunicationrelation}}\ \lin{}
\defcondition<zeroing>{zeroing}{\execution}\\
&&\qquad\qquad\attwf{\nonterminal{instrs}}(\execution)\wedge{}{\statesoftrace{\tau}_p}_{1}=\stateofprocess{\attat(\nonterminal{instrs})}{\infstamp{p}}{\LAMBDA{\areg{r}\in\setofallaregisters{p}}0}{\emptyset}
\nonumber
\end{eqntabular} 
\end{pcframed}

\caption{\LISA processes \label{fig:lisa-proc}}
\end{figure}

\FloatBarrier

\subsection{Lexems} 
For any process $p\in\interval{1}{n-1}$, where $n=\attpid(\nonterminal{prog})$ is the number of processes in the program $\nonterminal{prog}$, registers $\areg{r}\in\setofallaregisters{p}$ cannot be shared variable identifiers $\avar{x}\in\setofallavariables$, labels $l,\ell\in\setofallalabels{p}$ cannot be register or shared variable identifiers, and tags in tag sequences cannot be label, register, or
shared variable identifiers. This informal context condition is easy to include in the attribute grammar by collecting these sets and checking that their pairwise intersections are empty.

\subsection{Expressions} 
As shown in Fig.\nbsp\ref{fig:lisa-operations}, an \emph{operation} can be either an $\nonterminal{r-value}$, \ie a register or immediate value, or the result of an arithmetic (\eg\, \texttt{add, sub, mult}) or boolean (\eg\,  {\tt
eq, neq}) operator applied to a register ${{\areg{r}_2}}$ and an
$\nonterminal{r-value}$ ${\nonterminal{r-value}_3}$. The value $\evalexprenv[\areg{r}]$ of a register in process $p$ is
defined in next Sect.\nbsp\ref{sec:Local-sequentiality}.
\begin{figure}[!ht]
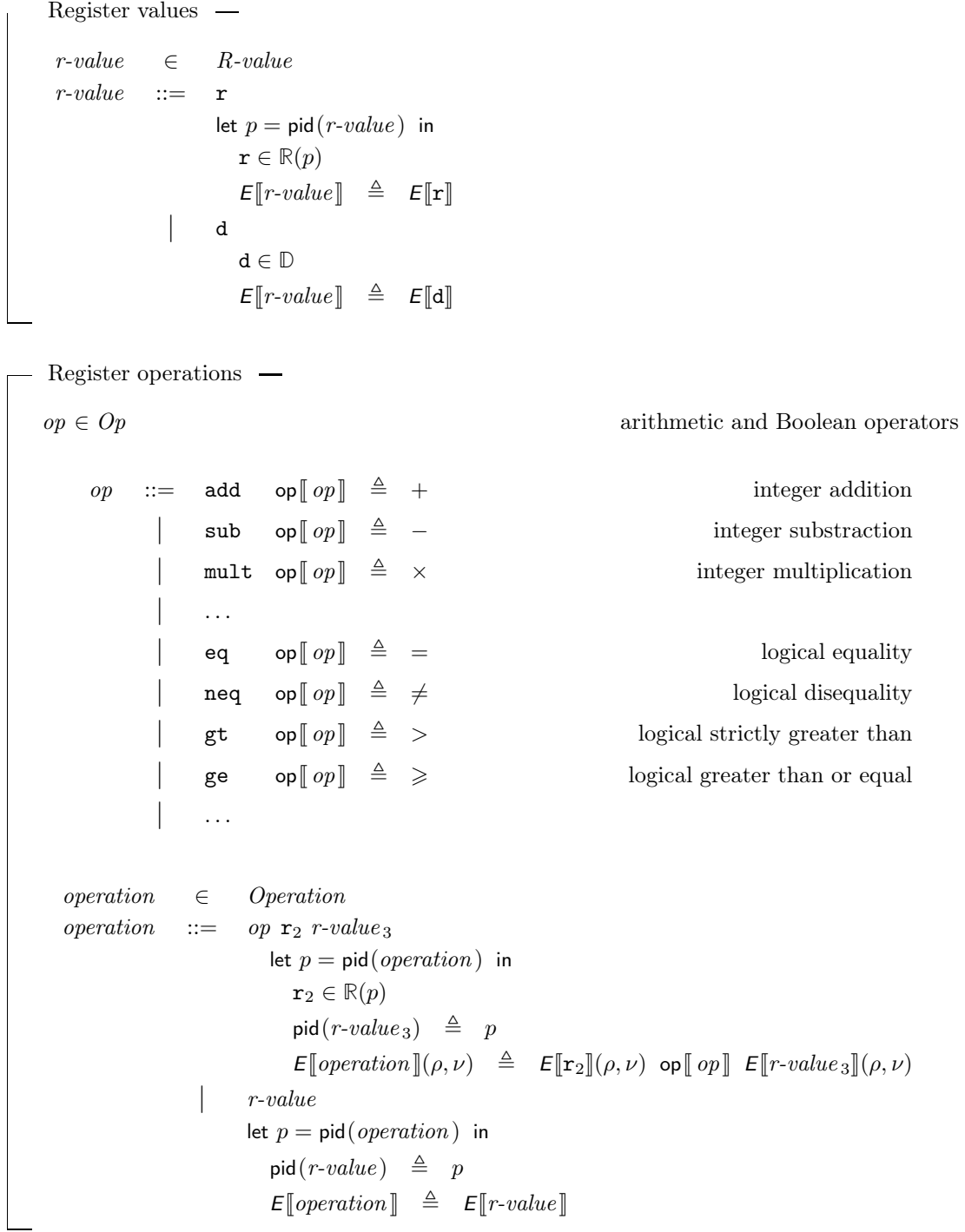


\begin{pcframed}[Register values]
\begin{eqntabular}[fl]{@{\quad}rcl}
\defnonterminal{r-value}&\in&\setofnonterminal{r-value}\nonumber\\
\nonterminal{r-value}&\grdef&{\areg{r}}\nonumber\\
&&\llet\ p = {\attpid(\nonterminal{r-value})}\ \lin\nonumber\\
&&\quad\areg{r}\in\setofallaregisters{p}\nonumber\\
&&\quad\evalexprenv[\nonterminal{r-value}]\colsep{\triangleq}\evalexprenv[\areg{r}]\nonumber\\
&\gror&\acst{d}\nonumber\\
&&\quad\acst{d}\in\defsetofalladenotations\nonumber\\
&&\quad\evalexprenv[\nonterminal{r-value}]\colsep{\triangleq}\evalexprenv[\acst{d}]\nonumber
\end{eqntabular}

\end{pcframed}

\begin{pcframed}[Register operations]
\medskip
\begin{description}\abovedisplayskip0pt
\item[\hskip1ex$\nonterminal{op}\in\setofnonterminal{op}$]\hfill arithmetic and Boolean operators\\[1ex]
\begin{eqntabular}[fl]{rcl@{\quad}lL}
\nonterminal{op}&\grdef{}&\oparith[add]{}{}&\attop{\nonterminal{op}}\colsep{\triangleq}{+}\renumber{integer addition\qquad\mbox{}}\\[0.5ex]
&{}\gror{}&\oparith[sub]{}{}&\attop{\nonterminal{op}}\colsep{\triangleq}{-}\renumber{integer substraction\qquad\mbox{}}\\[0.5ex]
&{}\gror{}&\oparith[mult]{}{}&\attop{\nonterminal{op}}\colsep{\triangleq}{\times}\renumber{integer multiplication\qquad\mbox{}}\\[0.5ex]
&{}\gror{}&\ldots\nonumber\\[0.5ex]
&{}\gror{}&\opbool[eq]{}{}&\attop{\nonterminal{op}}\colsep{\triangleq}{=}\renumber{logical equality\qquad\mbox{}}\\[0.5ex]
&{}\gror{}&\opbool[neq]{}{}&\attop{\nonterminal{op}}\colsep{\triangleq}{\neq}\renumber{logical disequality\qquad\mbox{}}\\[0.5ex]
&{}\gror{}&\opbool[gt]{}{}&\attop{\nonterminal{op}}\colsep{\triangleq}{>}\renumber{logical strictly greater than\qquad\mbox{}}\\[0.5ex]
&{}\gror{}&\opbool[ge]{}{}&\attop{\nonterminal{op}}\colsep{\triangleq}{\geqslant}\renumber{logical greater than or equal\qquad\mbox{}}\\[0.5ex]
&{}\gror{}&\ldots\nonumber
\end{eqntabular}
\end{description}

\begin{eqntabular}[fl]{@{\quad}rcl}
\
\defnonterminal{operation}&\in&\setofnonterminal{operation}\nonumber\\
\nonterminal{operation}&\grdef{}&\oparith[\nonterminal{op}]{{\areg{r}_2}}{\nonterminal{r-value}_3}\nonumber\\
&&\quad\llet\ p={\attpid(\nonterminal{operation})}\ \lin\nonumber\\
&&\qquad{\areg{r}_2}\in\setofallaregisters{p}\nonumber\\
&&\qquad\attpid({\nonterminal{r-value}_3})\colsep{\triangleq}p\nonumber\\
&&\qquad\evalexprenv[\nonterminal{operation}](\aenvironment,\avaluation)\colsep{\triangleq}{}\evalexprenv[\areg{r}_2](\aenvironment,\avaluation)\ \mathbin{\attop{\nonterminal{op}}}\ \evalexprenv[\nonterminal{r-value}_3](\aenvironment,\avaluation)\nonumber\\
&{}\gror{}&\nonterminal{r-value}\nonumber\\
&&\llet\ p={\attpid(\nonterminal{operation})}\ \lin\nonumber\\
&&\quad\attpid({\nonterminal{r-value}})\colsep{\triangleq}p\nonumber\\
&&\quad\evalexprenv[\nonterminal{operation}]\colsep{\triangleq}\evalexprenv[\nonterminal{r-value}]\nonumber
\end{eqntabular}
\end{pcframed}

\caption{Syntax of \LISA expressions \label{fig:lisa-operations}}
\end{figure}

\FloatBarrier

\subsection{Local sequentiality}\label{sec:Local-sequentiality}
The interleaved trace semantics of a process $\nonterminal{proc}$ of \LISA is \emph{locally sequential}. This means that
(1) the use of registers in a process are sequentially consistent in that the value of a local register $\areg{r}$ is its last 
assigned value and (2) that each process is executed in the process program order. 
Local sequentiality is much weaker than sequential consistency (SC) \cite{DBLP:journals/cacm/Keller76,DBLP:conf/mfcs/HennessyP79,DBLP:journals/tc/Lamport79} or sequential consistency per variable (SCPV) \cite{DBLP:journals/toplas/AlglaveMT14} which are relative to globally shared variables (and assume local sequentiality for local registers). 

\subsection{Events} 
All computation events collected in $\setofallevents\sqb{\nonterminal{prog}}$
have the form
$\mathfrak{e}({\markertuple{p}{\ell}{\nonterminal{u-instr}}{\theta}})$ or
$\mathfrak{e}({\markertuple{p}{\ell}{\nonterminal{u-instr}}{\theta}},{v})$
where $\mathfrak{e}$ is the name of the event ($\mathfrak{r}$ for read,
$\mathfrak{w}$ for write, \etc), $p$ is the process, and
${\nonterminal{u-instr}}$ is the instruction at label
${\ell}\in\setofallalabels{p}$ of the process that gave raise to that event.
The assignment $\mathfrak{a}$, read $\mathfrak{r}$, and write $\mathfrak{w}$
events may also carry a computed or communicated value $v\in\setofallintegers$.
Stamps $\theta$ are left unspecified but in case of an instruction
${\nonterminal{u-instr}}$ executed several times in a loop the can be used to
ensure that all the generated events are different (as required by the
\uniqueness condition \refcondition{uniqueness}{\semantics}). The conditions
$\attwf{\nonterminal{body}}$ make sure that the trace events record a program
execution. The traces may also include read-from events $\mathfrak{rf}$ which
are constraint by \refcondition{satisfaction}{\semantics} to
\refcondition{inception}{\semantics}.  It is checked in
Fig.\nbsp\ref{fig:lisa-progs} that the events on a trace must be generated by a
program execution and satisfy the constraints on communication event.

\subsection{Instructions} 

Instructions are given in Figure~\ref{fig:lisa-instr} and sequences of instructions in Figure~\ref{fig:lisa-instrs}. Except for the branch instructions, 
for which the label must be provided in the program, program instructions may be unlabelled.
In that case, an automatic labelling program transformation will add all missing labels. 
Therefore in the definition of a process $\nonterminal{proc}$ in Fig.\nbsp\ref{fig:lisa-procs}, all
instructions are assumed to have been labelled.

\begin{figure}[!ht]
\begin{pcframed}[Unlabelled instructions]
\begin{eqntabular*}[fl]{@{\quad}rcl}
\defnonterminal{u-instr}&\in&\setofnonterminal{u-instr}\\
\nonterminal{u-instr}&\grdef&\nonterminal{reg-u-instr}{}\gror{}\nonterminal{w-u-instr}{}\gror{}\nonterminal{r-u-instr}\gror{\nonterminal{rmw-u-instr}}\gror\nonterminal{b-u-instr}\\
&\gror&\nonterminal{f-u-instr}{}
\end{eqntabular*} 
For all these case $\nonterminal{u-instr}\grdef\ldots\gror\nonterminal{x-u-instr}\gror\ldots$, we have
\begin{eqntabular*}{l}
\attpid(\nonterminal{x-u-instr})\colsep{\triangleq}\attpid(\nonterminal{u-instr})\\
\setofallprocessidentifiers[\nonterminal{x-u-instr}]\colsep{\triangleq}\setofallprocessidentifiers[\nonterminal{u-instr}]\nonumber\\
\attinstr{\nonterminal{x-u-instr}}\colsep{\triangleq}\attinstr{\nonterminal{u-instr}}\nonumber\\
\setofallevents\sqb{\nonterminal{u-instrs}}\colsep{\triangleq}\setofallevents\sqb{\nonterminal{x-u-instr}}\\
\attat(\nonterminal{u-instr})\colsep{\triangleq}\attat(\nonterminal{x-u-instr})\\
\attafter(\nonterminal{x-u-instr})\colsep{\triangleq}\attafter(\nonterminal{u-instr})\\
\attwf{\nonterminal{u-instr}}\colsep{\triangleq}\attwf{\nonterminal{x-u-instr}}
\end{eqntabular*}
\end{pcframed}
\begin{pcframed}[Labelled instructions]
\begin{eqntabular*}[fl]{@{\quad}rcl@{\qquad}l}
\defnonterminal{instr}&\in&\setofnonterminal{instr}\nonumber\\
\nonterminal{instr}&\grdef&\phantom{{}\gror{}}{\grlabel}\colonsymbol\nonterminal{u-instr}&\attpid(\nonterminal{u-instr})\colsep{\triangleq}\attpid(\nonterminal{instr})\\
&&&\setofallprocessidentifiers[\nonterminal{u-instr}]\colsep{\triangleq}\setofallprocessidentifiers[\nonterminal{instr}]\nonumber\\
&&&\attinstr{\nonterminal{u-instr}}\colsep{\triangleq}\attinstr{\nonterminal{instr}}\nonumber\\
&&&\setofallevents\sqb{\nonterminal{instrs}}\colsep{\triangleq}\setofallevents\sqb{\nonterminal{u-instr}}\\
&          &                          &\attat(\nonterminal{instr})\colsep{\triangleq}\attat(\nonterminal{u-instr})\colsep{\triangleq}\grlabel\\
&          &                          &\attafter(\nonterminal{u-instr})\colsep{\triangleq}\attafter(\nonterminal{instr})\\
&&&\attwf{\nonterminal{instr}}\colsep{\triangleq}\attwf{\nonterminal{u-instr}}
\end{eqntabular*} 
\end{pcframed}
\caption{\LISA instructions \label{fig:lisa-instr}}
\end{figure}

\begin{figure}[!ht]
\begin{pcframed}[Sequences of labelled instructions]
\begin{eqntabular*}[fl]{@{\quad}rcl}
\defnonterminal{instrs}&\in&\setofnonterminal{instrs}\\
\nonterminal{instrs}&{}\grdef{}&\nonterminal{instr}\\
&&\qquad\quad\attpid(\nonterminal{instr})\colsep{\triangleq}\attpid(\nonterminal{instrs})\\
&&\qquad\quad\setofallprocessidentifiers[\nonterminal{instr}]\colsep{\triangleq}\setofallprocessidentifiers[\nonterminal{instrs}]\nonumber\\
&&\qquad\attinstr{\nonterminal{instrs}}\colsep{\triangleq}\attinstr{\nonterminal{instr}}\nonumber\\
&&\qquad\quad\setofallevents\sqb{\nonterminal{instrs}}\colsep{\triangleq}\setofallevents\sqb{\nonterminal{instr}}\\
&&\qquad\quad\attat(\nonterminal{instrs})\colsep{\triangleq}\attat(\nonterminal{instr})\\
&&\qquad\quad\attafter(\nonterminal{instr})\colsep{\triangleq}\attafter(\nonterminal{instrs})\\
&&\qquad\quad\attwf{\nonterminal{instrs}}\colsep{\triangleq}\attwf{\nonterminal{instr}}\\
&&{}\gror{}\nonterminal{instr}\mathrel{\semicolonsymbol}\nonterminal{instrs}_1\\
&&\qquad\quad\attpid(\nonterminal{instr})\colsep{\triangleq}\attpid(\nonterminal{instrs}_1)\colsep{\triangleq}\attpid(\nonterminal{instrs})\\
&&\qquad\quad\setofallprocessidentifiers[\nonterminal{instr}]\colsep{\triangleq}\setofallprocessidentifiers[\nonterminal{instrs}],\quad \setofallprocessidentifiers[\nonterminal{instrs}$_1$]\colsep{\triangleq}\setofallprocessidentifiers[\nonterminal{instrs}]\nonumber\\
&&\qquad\quad\attinstr{\nonterminal{instrs}}\colsep{\triangleq}\attinstr{\nonterminal{instr}}\cup\attinstr{\nonterminal{instrs}_1}\nonumber\\
&&\qquad\quad\setofallevents\sqb{\nonterminal{instrs}}\colsep{\triangleq}\setofallevents\sqb{\nonterminal{instr}}\cup\setofallevents\sqb{\nonterminal{instrs}_1}\\
&&\qquad\quad\attat(\nonterminal{instrs})\colsep{\triangleq}\attat(\nonterminal{instr})\\
&&\qquad\quad\attafter(\nonterminal{instr})\colsep{\triangleq}\attat(\nonterminal{instrs}_1)\\
&&\qquad\quad\attafter(\nonterminal{instrs}_1)\colsep{\triangleq}\attafter(\nonterminal{instr})\\
&&\qquad\quad\attwf{\nonterminal{instrs}}\colsep{\triangleq}\LAMBDA{\execution}\attwf{\nonterminal{instr}}\execution\wedge\attwf{\nonterminal{instrs}_1}\execution
\end{eqntabular*} 
\end{pcframed}
\caption{\LISA sequences of instructions \label{fig:lisa-instrs}}
\end{figure}
A special case $\nonterminal{reg-instrs}$ of sequences $\nonterminal{instrs}$ of instructions is considered 
in Fig.\nbsp\ref{fig:lisa-rmws} for the case where all instructions are register instructions in the read-modify-write instruction.

\FloatBarrier

\subsection{Markers}\label{sec:LISA:Markers}

\subsubsection{\bfseries $\bigl[$Labelled$\bigr]$ fences} are given in Figure~\ref{fig:lisa-fences}. Fences can only appear in processes (not in the program prelude or postlude).
Fences can have different names and can be labelled, in which case the labels must all occur in the same process as the fence. In this last case,
the fence is between any pair of actions with labels in the first and second set. Fences are just markers in the program and their  semantics is defined by the \cat semantics.
\begin{figure}[!ht]
\begin{pcframed}[{$\bigl[$Labelled$\bigr]$ fences}]
\begin{eqntabular}[fl]{@{\quad}rcl}
\defnonterminal{f-u-instr}&\in&\setofnonterminal{f-u-instr}\nonumber\\
\nonterminal{f-u-instr}&\grdef{}&
\finstra{\tagsequence}\ \bigl[\{\grlabel_1^0 \ldots \grlabel_1^m\}\ \{\grlabel_2^0 \ldots \grlabel_2^q\}\bigm],\quad m,q\geqslant 0
\nonumber\\
&&\llet\ p={\attpid(\nonterminal{f-u-instr})}\ \land\ \setofallprocessidentifiers=\setofallprocessidentifiers[\,\nonterminal{f-u-instr}]\ \land\ \ell={\attat(\nonterminal{f-u-instr})}\ \lin
\nonumber\\[1ex]
&&\quad\attinstr{\nonterminal{f-u-instr}}\colsep{\triangleq}\{\pair{\pair{p}{\ell}}{\finstra{\tagsequence}\ \bigl[\{\grlabel_1^0 \ldots \grlabel_1^m\}\ \{\grlabel_2^0 \ldots \grlabel_2^q\}\bigm]}\}\nonumber\\
&&\quad\setofallevents\sqb{\nonterminal{f-u-instr}}\colsep{\triangleq}\{\markerevent{\markertuple{p}{\ell}{\finstra{\tagsequence}\ \bigl[\{\grlabel_1^0 \ldots \grlabel_1^m\}\ \{\grlabel_2^0 \ldots \grlabel_2^q\}\bigr]}{\theta}}\mid\theta\in\setofallstamps{p}\}\nonumber\\[1ex]
&&\quad\attwf{\nonterminal{f-u-instr}}\colsep{\triangleq}\LAMBDA{\execution}
\llet\ \execution={\parallelexecution{\trace_{\startprocess}}{p\in\setofallprocessidentifiers}{}{\trace_p}{\trace_{\finishprocess}}{\rfcommunicationrelation}}\lin
\defcondition<fence>{fence}{\execution}\\
&&\qquad
\forall k\in\interval[open left, open right]{1}{1+\lengthoftrace{\tau_p}}\suchthat
\forall\acontrol'\in\setofallalabels{p}\suchthat
\forall\aenvironment,\aenvironment'\in\setofallaenvironments{p}\suchthat\nonumber\\
&&\qquad\quad
\forall\avaluation,\avaluation'\in\setofallavaluations{p}\suchthat
\forall\theta,\theta'\in\setofallstamps{p}\suchthat\nonumber\\
&&\qquad\qquad({\statesoftrace{\tau_p}}_{k-1}=\stateofprocess{\ell}{\theta}{\aenvironment}{\avaluation}\wedge{}\nonumber\\
&&\qquad\qquad\phantom{(}{\eventsoftrace{\tau_p}}_{k}=\markerevent{\markertuple{p}{\ell}{\finstra{\tagsequence}\ \bigl[\{\grlabel_1^0 \ldots \grlabel_1^m\}\ \{\grlabel_2^0 \ldots \grlabel_2^q\}\bigr]}{\theta}}
\wedge{}\nonumber\\
&&\qquad\qquad\phantom{(}{\statesoftrace{\tau_p}}_{k}=\stateofprocess{\acontrol'}{\theta'}{\aenvironment'}{\avaluation'}){}\implies{}\nonumber\\
&&\qquad\qquad\qquad(\acontrol'={\attafter(\nonterminal{f-u-instr})}\wedge\theta'=\succstamp{p}(\theta)\wedge\aenvironment'=\aenvironment\wedge\avaluation'=\avaluation)\ .\nonumber
\end{eqntabular}
($\bigl[\ldots\bigr]$ indicates that $\ldots$ is optional)
\end{pcframed}
\caption{\LISA fences \label{fig:lisa-fences}}
\end{figure}

\subsubsection{\bfseries {\normalfont$\ttfont{rmw}$} delimiters} $\grbeginrmw[\tagsequence]{x}$ and $\grendrmw[\tagsequence]{x}$ in Fig.\nbsp\ref{fig:lisa-rmws} are markers used to
delimit read, modify, and write instructions $\ttfont{rmw}$ which atomically update shared variable variable $\avar{x}$ as defined in Sect.\nbsp\ref{sec:lisa-rmw}. The fact that read, modify, and write instructions should be atomic will follow from the definition of their semantics in a \cat specification.

\subsection{Actions}\label{sec:LISA:Actions}

\subsubsection{\bfseries Register instructions} are given in
Fig.\nbsp\ref{fig:lisa-regs} where $\environmentassignment{f}{x}{v}(x)=v$ and $\environmentassignment{f}{x}{v}(y)=f(y)$ when $y\neq x$.  \LISA register accesses are of the form
$\mova{{\areg{r}_1}}{\nonterminal{operation}}$.
Namely, they \emph{move} the result of an \emph{operation}
into a register, \eg ${{\areg{r}_1}}$.

\begin{figure}[!ht]
\begin{pcframed}[Register instructions]
\begin{eqntabular}[fl]{@{\quad}rcl}
\defnonterminal{reg-u-instr}&\in&\setofnonterminal{reg-u-instr}\nonumber\\
\nonterminal{reg-u-instr}&{}\grdef{}&\mova{{\areg{r}_1}}{\nonterminal{operation}}\nonumber\\
&&\quad\llet\ p={\attpid(\nonterminal{reg-u-instr})}\ \land\ \setofallprocessidentifiers=\setofallprocessidentifiers[\,\nonterminal{reg-u-instr}]\ \land\ \ell={\attat(\nonterminal{reg-u-instr})}\ \lin
\nonumber\\[1ex]
&&\qquad{\areg{r}_1}\in\setofallaregisters{p}\nonumber\\[1ex]
&&\qquad\attpid({\nonterminal{operation}})\colsep{\triangleq}p\nonumber\\[1ex]
&&\qquad\attinstr{\nonterminal{reg-u-instr}}\colsep{\triangleq}\{\pair{\pair{p}{\ell}}{\mova{{\areg{r}_1}}{\nonterminal{operation}}}\}\nonumber\\
&&\qquad\setofallevents\sqb{\nonterminal{reg-u-instr}}\colsep{\triangleq}{}\{\assignmentevent{\markertuple{p}{\ell}{\mova{{\areg{r}_1}}{\nonterminal{operation}}}{\theta}}{v}\mid\theta\in\setofallstamps{p}\wedge v\in\setofallintegers\}\nonumber\\[1ex]
&&\qquad\attwf{\nonterminal{reg-u-instr}}\colsep{\triangleq}
\defcondition<assignment>{assignment}{\execution}\\
&&\qquad\quad\LAMBDA{\execution}\llet\ \execution ={\parallelexecution{\trace_{\startprocess}}{p\in\setofallprocessidentifiers}{}{\trace_p}{\trace_{\finishprocess}}{\rfcommunicationrelation}}\ \lin\nonumber\\
&&\qquad\qquad
\forall k\in\interval[open left, open right]{1}{1+\lengthoftrace{\tau_p}}\suchthat
\forall\acontrol'\in\setofallalabels{p}\suchthat
\forall\aenvironment,\aenvironment'\in\setofallaenvironments{p}\suchthat\nonumber\\
&&\quad\qquad\qquad
\forall\avaluation,\avaluation'\in\setofallavaluations{p}\suchthat
\forall\theta,\theta'\in\setofallstamps{p}\suchthat
\forall v\in\setofallintegers\suchthat\nonumber\\
&&\qquad\qquad\qquad({\statesoftrace{\tau_p}}_{k-1}=\stateofprocess{\ell}{\theta}{\aenvironment}{\avaluation}\wedge{}\nonumber\\
&&\qquad\qquad\qquad\phantom{)}{\eventsoftrace{\tau_p}}_{k}=\assignmentevent{\markertuple{p}{\ell}{\mova{{\areg{r}_1}}{\nonterminal{operation}}}{\theta}}{v}\wedge{}\nonumber\\
&&\qquad\qquad\qquad\phantom{)}{\statesoftrace{\tau_p}}_{k}=\stateofprocess{\acontrol'}{\theta'}{\aenvironment'}{\avaluation'}){}\implies{}\nonumber\\
&&\qquad\qquad\qquad\qquad{}(v=\evalexprenv[{\nonterminal{operation}}](\aenvironment,\avaluation)\wedge{}\acontrol'={}\attafter(\nonterminal{reg-u-instr}){}\wedge{}\nonumber\\
&&\qquad\qquad\qquad\qquad{}\phantom{(}\theta'=\succstamp{p}(\theta)\wedge\aenvironment'=\environmentassignment{\aenvironment}{\areg{r}_1}{v}\wedge{}\avaluation'=\avaluation)\ .\nonumber
\end{eqntabular}
\end{pcframed}

\caption{Syntax of \LISA register accesses \label{fig:lisa-regs}}
\end{figure}

\subsubsection{\bfseries Read.}\quad
In a read instruction of Fig~\ref{fig:lisa-reads}, the
$\nonterminal{r-value}$ denotes the value assigned to register $\areg{r}$.

\subsubsection{\bfseries Write.}\quad
Write accesses are given in Figure~\ref{fig:lisa-writes}.  The value written is that of ${\nonterminal{r-value}}$.

\subsubsection{\bfseries Read, modify, and write.}\quad\label{sec:lisa-rmw}
Read-modify-write accesses are given in Figure~\ref{fig:lisa-rmws}. Read-modify-write instructions
$\rmw[\tagsequence]{\areg{r}}{\nonterminal{reg-instrs}}{\avar{x}}{}{}$
can only appear in processes (not in the program prelude or postlude). The sequence of  register
accesses $\nonterminal{reg-instrs}$ is defined in Fig.\nbsp\ref{fig:lisa-regs} as a sequence of (labelled or appropriately labelled by fresh labels) register accesses.

In a read-modify-write instruction
$\rmw[\tagsequence]{\areg{r}}{\nonterminal{reg-instrs}}{\avar{x}}{\nonterminal{r-value}}{}$,
the last of the register instructions in $\nonterminal{reg-instrs}$ should
assign a value to register \areg{r}.

The \texttt{rmw} instructions
\begin{eqntabular*}{c}
\rmw[\tagsequence]{\areg{r}}{\nonterminal{reg-instrs}}{\avar{x}}{}{}
\end{eqntabular*}
are compiled into a sequence of concrete instructions delimited by $\grbeginrmw[\tagsequence]{x}$ and $\grendrmw[\tagsequence]{x}$ markers  as follows:
\begin{eqntabular*}[fl]{@{\quad}l}
\grbeginrmw[\tagsequence]{x}\semicolonsymbol\\
\rinstra[\tagsequence]{{\areg{r}}}{\avar{x}}{}\semicolonsymbol\\
\nonterminal{reg-instrs}\semicolonsymbol\\
\winstra[\tagsequence]{\avar{x}}{}{{\areg{r}}}\semicolonsymbol\\
\grendrmw[\tagsequence]{x}
\end{eqntabular*} 
before being parsed by the attribute grammar (which therefore contains no $\rmw[\tagsequence]{\areg{r}}{\nonterminal{reg-instrs}}{\avar{x}}{}{}$ instrcution). The last register instruction in $\nonterminal{reg-instrs}$ must be an assignment to register ${\areg{r}}$. A \texttt{cat} specification must be used to specify
atomicity. When abstracting the traces to candidate executions for \cat in Sect.\nbsp\ref{sec:abstraction-to-candidate-execution},
the abstraction $\alphacandidateexecution(\tau)$ will replace the two consecutive events $\markerevent{\markertuple{p}{\ell_1}{\grbeginrmw[\tagsequence]{x}}{\theta_1}}
\readevent{\markertuple{p}{\ell_2}{\rinstra[\tagsequence]{{\areg{r}}}{\avar{x}}{}}{\theta_2}}{v}$ on $\tau$ by $\readevent{\markertuple{p}{\ell_2}{\rinstraast[\tagsequence]{{\areg{r}}}{\avar{x}}{}}{\theta_2}}{v}$ and $\writeevent{\markertuple{p}{\ell_1}{\winstra[\tagsequence]{\avar{x}}{}{{\areg{r}}}}{\theta_1}}{v}
\markerevent{\markertuple{p}{\ell_2}{\grendrmw[\tagsequence]{x}}{\theta_2}}$ by $\writeevent{\markertuple{p}{\ell_1}{\winstraast[\tagsequence]{\avar{x}}{}{{\areg{r}}}}{\theta_1}}{v}$ to conform to $\cat$ conventions \cite{AlglaveCousotMaranget-cat-HSA-2015}.

\begin{figure}[!ht]
\begin{pcframed}[Write accesses]
\begin{eqntabular}[fl]{@{\quad}rcl}
\defnonterminal{w-u-instr}&\in&\setofnonterminal{w-u-instr}\nonumber\\
\nonterminal{w-u-instr}&\grdef{}&\winstra[\tagsequence]{\avar{x}}{}{\nonterminal{r-value}}\nonumber\\
&&\quad\llet\ p={\attpid(\nonterminal{w-u-instr})}\ \land\ \setofallprocessidentifiers=\setofallprocessidentifiers[\,\nonterminal{w-u-instr}]\nonumber\\
&&\qquad\land\ \ell={\attat(\nonterminal{w-u-instr})}\ \lin\nonumber\\
&&\qquad\quad\attpid({\nonterminal{r-value}})\colsep{\triangleq}p\nonumber\\[1ex]
&&\qquad\quad\attinstr{\nonterminal{w-u-instr}}\colsep{\triangleq}\{\pair{\pair{p}{\ell}}{\winstra[\tagsequence]{\avar{x}}{}{\nonterminal{r-value}}}\}\nonumber\\[1ex]
&&\qquad\quad\setofallevents\sqb{\nonterminal{w-u-instr}}\colsep{\triangleq}\{{\writeevent{\markertuple{p}{\ell}{\winstra[\tagsequence]{\avar{x}}{}{\nonterminal{r-value}}}{\theta}}{v}}\mid\theta\in\setofallstamps{p}\wedge v\in\setofallintegers\}\nonumber\\[1ex]
&&\qquad\quad\attwf{\nonterminal{w-u-instr}}\colsep{\triangleq}
\defcondition<awrite>{awrite}{\execution}\\
&&\qquad\qquad\LAMBDA{\execution}\llet\ \execution=
{\parallelexecution{\trace_{\startprocess}}{p\in\setofallprocessidentifiers}{}{\trace_p}{\trace_{\finishprocess}}{\rfcommunicationrelation}}\ \lin\nonumber\\
&&\qquad\qquad\quad\forall k\in\interval[open left, open right]{1}{1+\lengthoftrace{\tau_p}}\suchthat
\forall\acontrol'\in\setofallalabels{p}\suchthat
\forall\aenvironment,\aenvironment'\in\setofallaenvironments{p}\suchthat\nonumber\\
&&\qquad\qquad\qquad
\forall\avaluation,\avaluation'\in\setofallavaluations{p}\suchthat
\forall\theta,\theta'\in\setofallstamps{p}\suchthat
\forall v\in\setofallintegers\suchthat\nonumber\\
&&\qquad\qquad\qquad\quad({\statesoftrace{\tau_p}}_{k-1}=\stateofprocess{\ell}{\theta}{\aenvironment}{\avaluation}\wedge{}\nonumber\\
&&\qquad\qquad\qquad\quad\phantom{(}{\eventsoftrace{\tau_p}}_{k}=\writeevent{\markertuple{p}{\ell}{\winstra[\tagsequence]{\avar{x}}{}{\nonterminal{r-value}}}{\theta}}{v}\wedge{}\nonumber\\
&&\qquad\qquad\qquad\quad\phantom{(}{\statesoftrace{\tau_p}}_{k}=\stateofprocess{\acontrol'}{\theta'}{\aenvironment'}{\avaluation'})
\implies{}\nonumber\\
&&\qquad\qquad\qquad\qquad(v=\evalexprenv[{\nonterminal{r-value}}](\aenvironment,\avaluation)\wedge{}\acontrol'={}\attafter(\nonterminal{w-u-instr}){}\wedge{}\nonumber\\
&&\qquad\qquad\qquad\qquad{}\phantom{(}\theta'=\succstamp{p}(\theta)\wedge\aenvironment=\aenvironment'\wedge{}\avaluation'=\avaluation)\ .\nonumber
\end{eqntabular}
(this includes the initialization writes for $p=\startprocess$.)
\end{pcframed}

\caption{\LISA memory write accesses  \label{fig:lisa-writes}}
\end{figure}

\begin{figure}[!ht]
\begin{pcframed}[Read accesses]
\begin{eqntabular}[fl]{@{\quad}rcl}
\defnonterminal{r-u-instr}&\in&\setofnonterminal{r-u-instr}\nonumber\\
\nonterminal{r-u-instr}&\grdef{}&
\rinstra[\tagsequence]{{\areg{r}_1}}{\avar{x}}{}\nonumber\\[1ex]
&&\quad\llet\ p={\attpid(\nonterminal{r-u-instr})}\ \land\ \setofallprocessidentifiers=\setofallprocessidentifiers[\,\nonterminal{r-u-instr}]\nonumber\\
&&\qquad \land\ \ell={\attat(\nonterminal{r-u-instr})}\ \lin\nonumber\\
&&\qquad\quad{\areg{r}_1}\in\setofallaregisters{p}\nonumber\\[1ex]
&&\qquad\quad\attinstr{\nonterminal{r-u-instr}}\colsep{\triangleq}\{\pair{\pair{p}{\ell}}{\rinstra[\tagsequence]{{\areg{r}_1}}{\avar{x}}{}}\}\nonumber\\[1ex]
&&\qquad\quad\setofallevents\sqb{\nonterminal{r-u-instr}}\colsep{\triangleq}\{{\readevent{\markertuple{p}{\ell}{\rinstra[\tagsequence]{{\areg{r}_1}}{\avar{x}}{}}{\theta}}{\svar{x}{\theta}}}\mid\theta\in\setofallstamps{p}\}\nonumber\\[1ex]
&&\qquad\quad\attwf{\nonterminal{r-u-instr}}\colsep{\triangleq}
\defcondition<aread>{aread}{\execution}\\
&&\qquad\qquad\LAMBDA{\execution}
\llet\ \execution={\parallelexecution{\trace_{\startprocess}}{p\in\setofallprocessidentifiers}{}{\trace_p}{\trace_{\finishprocess}}{\rfcommunicationrelation}}\lin\nonumber\\
&&\qquad\qquad\quad
\forall k\in\interval[open left, open right]{1}{1+\lengthoftrace{\tau_p}}\suchthat
\forall\acontrol'\in\setofallalabels{p}\suchthat
\forall\aenvironment,\aenvironment'\in\setofallaenvironments{p}\suchthat\nonumber\\
&&\qquad\qquad\qquad
\forall\avaluation,\avaluation'\in\setofallavaluations{p}\suchthat
\forall\theta,\theta'\in\setofallstamps{p}\suchthat
\forall v\in\setofallintegers\suchthat\nonumber\\
&&\qquad\qquad\qquad\quad({\statesoftrace{\tau_p}}_{k-1}=\stateofprocess{\ell}{\theta}{\aenvironment}{\avaluation}\wedge{}\nonumber\\
&&\qquad\qquad\qquad\quad\phantom{(}{\eventsoftrace{\tau_p}}_{k}=\readevent{\markertuple{p}{\ell}{\rinstra[\tagsequence]{{\areg{r}_1}}{\avar{x}}{}}{\theta}}{\svar{x}{\theta}}\wedge{}\nonumber\\
&&\qquad\qquad\qquad\quad\phantom{(}{\statesoftrace{\tau_p}}_{k}=\stateofprocess{\acontrol'}{\theta'}{\aenvironment'}{\avaluation'})\implies{}\nonumber\\
&&\qquad\qquad\qquad\qquad(\acontrol'={}\attafter(\nonterminal{r-u-instr}){}\wedge{}\theta'=\succstamp{p}(\theta)\wedge{}\aenvironment'=\environmentassignment{\aenvironment}{{\areg{r}_1}}{\svar{x}{\theta}}{}\wedge{}\nonumber\\
&&\qquad\qquad\qquad\qquad{}\phantom{(}\exists{q\in\setofallprocessidentifiersstart}\suchthat\exists! j\in\interval[open right]{1}{1+\lengthoftrace{\tau_q}}\suchthat\exists \ell', \theta'',v\suchthat\nonumber\\
&&\qquad\qquad\qquad\qquad\quad{}\phantom{(}({\eventsoftrace{\tau_q}}_{j}={\writeevent{\markertuple{q}{\ell'}{\winstra[\tagsequence]{\avar{x}}{}{\nonterminal{r-value}}}{\theta''}}{v}}\wedge{}\nonumber\\
&&\qquad\qquad\qquad\qquad\quad{}\phantom{((}
\catrfevent{{\eventsoftrace{\tau_q}}_{j}}{{\eventsoftrace{\tau_p}}_{k}}\in\rfcommunicationrelation\wedge{}
\avaluation'=\environmentassignment{\avaluation}{\svar{x}{\theta}}{v}))\ .\nonumber
\end{eqntabular}
(this includes the finalization reads for $p=\finishprocess$, in which case ${{\areg{r}_1}}={{\areg{r}_{\avar{x}}}}$).
\end{pcframed}

\caption{\LISA memory read accesses \label{fig:lisa-reads}}
\end{figure}

\begin{figure}[!ht]
\begin{pcframed}[Sequences of labelled register instructions]
\begin{eqntabular*}[fl]{@{\quad}rcl}
\defnonterminal{reg-instrs}&\in&\setofnonterminal{reg-instrs}\\
\nonterminal{reg-instrs}&{}\grdef{}&{\grlabel}\colonsymbol\nonterminal{reg-u-instr}\\
&&\qquad\attpid(\nonterminal{reg-u-instr})\colsep{\triangleq}{\attpid(\nonterminal{reg-instrs})}\\
&&\qquad\setofallprocessidentifiers[\nonterminal{reg-u-instr}]\colsep{\triangleq}\setofallprocessidentifiers[\nonterminal{reg-instrs}]\nonumber\\
&&\qquad\attinstr{\nonterminal{reg-instrs}}\colsep{\triangleq}\attinstr{\nonterminal{reg-u-instr}}\nonumber\\
&&\qquad\setofallevents\sqb{\nonterminal{reg-u-instrs}}\colsep{\triangleq}\setofallevents\sqb{\nonterminal{reg-u-instr}}\\
&&\qquad\attat(\nonterminal{reg-instrs})\colsep{\triangleq}\attat(\nonterminal{reg-instr})\colsep{\triangleq}{\grlabel}\\
&&\qquad\attafter(\nonterminal{reg-u-instr})\colsep{\triangleq}\attafter(\nonterminal{reg-instrs})\\
&&\qquad\attwf{\nonterminal{reg-instrs}}\colsep{\triangleq}\attwf{\nonterminal{reg-u-instr}}\\
&{}\gror{} &\nonterminal{reg-instr}\mathrel{\semicolonsymbol}\nonterminal{reg-instrs}_1\\
&&\qquad\attpid(\nonterminal{reg-instr})\colsep{\triangleq}\attpid(\nonterminal{reg-instrs}_1)\colsep{\triangleq}\attpid(\nonterminal{reg-instrs})\\
&&\qquad\setofallprocessidentifiers[\nonterminal{reg-instr}]\colsep{\triangleq}\setofallprocessidentifiers[\nonterminal{reg-instrs}$_1$]\colsep{\triangleq}\setofallprocessidentifiers[\nonterminal{reg-instrs}]\nonumber\\
&&\qquad\attinstr{\nonterminal{reg-instrs}}\colsep{\triangleq}\attinstr{\nonterminal{reg-u-instr}}\cup\attinstr{\nonterminal{reg-instrs}_1}\nonumber\\
&&\qquad\setofallevents\sqb{\nonterminal{reg-u-instrs}}\colsep{\triangleq}\setofallevents\sqb{\nonterminal{reg-u-instr}}\cup\setofallevents\sqb{\nonterminal{reg-u-instrs}_1}\\
&&\qquad\attat(\nonterminal{reg-instrs})\colsep{\triangleq}\attat(\nonterminal{reg-instr})\\
&&\qquad\attafter(\nonterminal{reg-instr})\colsep{\triangleq}\attat(\nonterminal{reg-instrs}_1)\\
&&\qquad\attafter(\nonterminal{reg-instrs}_1)\colsep{\triangleq}\attafter(\nonterminal{reg-instrs})\\
&&\qquad\attwf{\nonterminal{reg-instrs}}\colsep{\triangleq}\LAMBDA{\execution}\attwf{\nonterminal{reg-instr}}(\execution)\wedge\attwf{\nonterminal{reg-instrs}_1}(\execution)\end{eqntabular*}
\end{pcframed}

\caption{\LISA sequences of labelled register instructions \label{fig:lisa-reg-instrs}}
\end{figure}

\begin{figure}[!ht]
\begin{pcframed}[Read-modify-write accesses]
\begin{eqntabular}[fl]{@{\quad}rcl}
\ \defnonterminal{rmw-u-instr}&\in&\setofnonterminal{rmw-u-instr}\quad \renumber{(originating form $\rmw[\tagsequence]{\areg{r}}{\nonterminal{reg-instrs}}{\avar{x}}{\nonterminal{r-value}}$)}\\
\nonterminal{rmw-u-instr}&\grdef&\grbeginrmw[\tagsequence]{x}\nonumber\\
&&\quad\llet\ p={\attpid(\nonterminal{rmw-u-instr})}\ \land\ \setofallprocessidentifiers=\setofallprocessidentifiers[\,\nonterminal{rmw-u-instr}]\nonumber\\
&&\qquad\land\ \ell={\attat(\nonterminal{rmw-u-instr})}\ \lin\nonumber\\
&&\qquad\quad\attinstr{\nonterminal{rmw-u-instr}}\colsep{\triangleq}\{\pair{\pair{p}{\ell}}{\grbeginrmw[\tagsequence]{x}}\}\nonumber\\
&&\qquad\quad\setofallevents\sqb{\nonterminal{rmw-u-instr}}\colsep{\triangleq}\{{\markerevent{\markertuple{p}{\ell}{\grbeginrmw[\tagsequence]{x}}{\theta}}}\mid\theta\in\setofallstamps{p}\}\nonumber\\
&&\qquad\quad\attwf{\nonterminal{rmw-u-instr}}\colsep{\triangleq}
\defcondition<beginrmw>{beginrmw}[b]{\execution}\\
&&\qquad\qquad\LAMBDA{\execution}\llet\ \execution={\parallelexecution{\trace_{\startprocess}}{p\in\setofallprocessidentifiers}{}{\trace_p}{\trace_{\finishprocess}}{\rfcommunicationrelation}}\ \lin\nonumber\\
&&\qquad\qquad\quad\forall k\in\interval[open right]{1}{1+\lengthoftrace{\tau_p}}\suchthat\forall\theta\in\setofallstamps{p}\suchthat
\forall\acontrol'\in\setofallalabels{p}\suchthat
\forall\aenvironment,\aenvironment'\in\setofallaenvironments{p}\suchthat\nonumber\\
&&\qquad\qquad\quad\forall\avaluation,\avaluation'\in\setofallavaluations{p}\suchthat
\forall\theta,\theta'\in\setofallstamps{p}\suchthat\nonumber\\
&&\qquad\qquad\qquad({\statesoftrace{\tau_p}}_{k-1}=\stateofprocess{\ell}{\theta}{\aenvironment}{\avaluation}\wedge{}\nonumber\\
&&\qquad\qquad\qquad\phantom{(}{\eventsoftrace{\tau_p}}_{k}=\markerevent{\markertuple{p}{\ell}{\grbeginrmw[\tagsequence]{x}}{\theta}}{}\wedge{}\nonumber\\
&&\qquad\qquad\qquad\phantom{(}{\statesoftrace{\tau_p}}_{k}=\stateofprocess{\acontrol'}{\theta'}{\aenvironment'}{\avaluation'}){}\implies{}\nonumber\\
&&\qquad\qquad\qquad\qquad{}(\acontrol'=\attafter(\nonterminal{rmw-u-instr}){}\wedge{}
\theta'=\succstamp{p}(\theta)\wedge{}
\aenvironment'=\aenvironment\wedge{}
\avaluation'=\avaluation)\ .\nonumber
\\[1ex]
&\gror& \grendrmw[\tagsequence]{x}\nonumber\\
&&\quad\llet\ p={\attpid(\nonterminal{rmw-u-instr})}\ \land\ \setofallprocessidentifiers=\setofallprocessidentifiers[\,\nonterminal{rmw-u-instr}]\nonumber\\
&&\qquad \land\ \ell={\attat(\nonterminal{rmw-u-instr})}\ \lin\nonumber\\
&&\qquad\quad\attinstr{\nonterminal{rmw-u-instr}}\colsep{\triangleq}\{\pair{\pair{p}{\ell}}{\grendrmw[\tagsequence]{x}}\}\nonumber\\
&&\qquad\quad\setofallevents\sqb{\nonterminal{rmw-u-instr}}\colsep{\triangleq}\{{\markerevent{\markertuple{p}{\ell}{\grendrmw[\tagsequence]{x}}{\theta}}}\mid\theta\in\setofallstamps{p}\}\nonumber\\
&&\qquad\quad\attwf{\nonterminal{rmw-u-instr}}\colsep{\triangleq}
\renumber{\refcondition{beginrmw}[e]{\execution}}\\
&&\qquad\qquad\LAMBDA{\execution}{\parallelexecution{\trace_{\startprocess}}{p\in\setofallprocessidentifiers}{}{\trace_p}{\trace_{\finishprocess}}{\rfcommunicationrelation}}
\nonumber\\
&&\qquad\qquad\quad\forall k\in\interval[open right]{1}{1+\lengthoftrace{\tau}}\suchthat\forall\theta\in\setofallstamps{p}\suchthat
\forall\acontrol'\in\setofallalabels{p}\suchthat
\forall\aenvironment,\aenvironment'\in\setofallaenvironments{p}\suchthat\nonumber\\
&&\qquad\qquad\quad\forall\avaluation,\avaluation'\in\setofallavaluations{p}\suchthat
\forall\theta,\theta'\in\setofallstamps{p}\suchthat\nonumber\\
&&\qquad\qquad\qquad({\statesoftrace{\tau_p}}_{k-1}=\stateofprocess{\ell}{\theta}{\aenvironment}{\avaluation}\wedge{}\nonumber\\
&&\qquad\qquad\qquad\phantom{(}{\eventsoftrace{\tau_p}}_{k}=\markerevent{\markertuple{p}{\ell}{\grendrmw[\tagsequence]{x}}{\theta}}\wedge\nonumber\\
&&\qquad\qquad\qquad\phantom{(}{\statesoftrace{\tau_p}}_{k}=\stateofprocess{\acontrol'}{\theta'}{\aenvironment'}{\avaluation'}){}\implies{}\nonumber\\
&&\qquad\qquad\qquad\qquad{}({}\acontrol'={}\attafter(\nonterminal{rmw-u-instr}){}\wedge{}\theta'=\succstamp{p}(\theta)\wedge{}\aenvironment'=\aenvironment\wedge{}\avaluation'=\avaluation)\ .\nonumber
\end{eqntabular} 
\end{pcframed}

\caption{\LISA memory read-modify-write accesses \label{fig:lisa-rmws}}
\end{figure}

\subsubsection{\bfseries Branch.}\quad

Branches are given in Figure~\ref{fig:lisa-branches}. 
\begin{figure}[!ht]
\begin{pcframed}[Branches]
\begin{eqntabular}[fl]{@{\quad}rcl}
\defnonterminal{b-u-instr}&\in&\setofnonterminal{b-u-instr}\nonumber\\
\nonterminal{b-u-instr}&\grdef&\binstra[\tagsequence]{\nonterminal{operation}}{\grlabel_{\mskip-2mu t}}\nonumber\\
&&\quad\llet\ p={\attpid(\nonterminal{b-u-instr})}\ \land\ \setofallprocessidentifiers=\setofallprocessidentifiers[\,\nonterminal{b-u-instr}]\ \land\ \ell={\attat(\nonterminal{b-u-instr})}\ \lin\nonumber\\
&&\qquad{\attpid(\nonterminal{operation})}\colsep{\triangleq}p\nonumber\\
&&\qquad{\attinstr{\nonterminal{b-u-instr}}}\colsep{\triangleq}\{\pair{\pair{p}{\ell}}{\binstra[\tagsequence]{\nonterminal{operation}}{\grlabel_{\mskip-2mu t}}}\}\nonumber\\
&&\qquad\setofallevents\sqb{\nonterminal{b-u-instr}}\colsep{\triangleq}\{\begin{array}[t]{@{}l@{}}
{\testevent{\markertuple{p}{\ell}{\binstra[\tagsequence]{\nonterminal{operation}}{\grlabel_{\mskip-2mu t}}}{\theta}}){}},{}\nonumber\\
{\testevent{\markertuple{p}{\ell}{\binstra[\tagsequence]{\mneg \nonterminal{operation}}{\grlabel_{\mskip-2mu t}}}{\theta}})}\mid\theta\in\setofallstamps{p}\}\end{array}\nonumber\\
&&\qquad\attwf{\nonterminal{b-u-instr}}\colsep{\triangleq}
\nonumber\\
&&\qquad\quad
\LAMBDA{\execution}\llet\ \execution={\parallelexecution{\trace_{\startprocess}}{p\in\setofallprocessidentifiers}{}{\trace_p}{\trace_{\finishprocess}}{\rfcommunicationrelation}}\ \lin
\nonumber\\
&&\qquad\qquad\forall k\in\interval[open right]{1}{1+\lengthoftrace{\tau_p}}\suchthat
\forall\acontrol'\in\setofallalabels{p}\suchthat
\forall\theta,\theta'\in\setofallstamps{p}\suchthat\forall\aenvironment,\aenvironment'\in\setofallaenvironments{p}\suchthat\nonumber\\
&&\qquad\qquad\forall\avaluation,\avaluation'\in\setofallavaluations{p}\suchthat
\nonumber\\
&&\qquad\qquad\quad({\statesoftrace{\tau_p}}_{k-1}=\stateofprocess{\ell}{\theta}{\aenvironment}{\avaluation}\wedge{}\defcondition<test>{test}[t]{\execution}\\
&&\qquad\qquad\quad\phantom{(}{\eventsoftrace{\tau_p}}_{k}=\testevent{\markertuple{p}{\ell}{\binstra[\tagsequence]{\nonterminal{operation}}{\grlabel_{\mskip-2mu t}}{}}{\theta}}{}\wedge{}\nonumber\\
&&\qquad\qquad\quad\phantom{(}{\statesoftrace{\tau_p}}_{k}=\stateofprocess{\acontrol'}{\theta'}{\aenvironment'}{\avaluation'}){}\implies{}\nonumber\\
&&\qquad\qquad\qquad{}(\satisfiable{\evalexprenv[\nonterminal{operation}](\aenvironment,\avaluation)\neq0}\wedge{}\acontrol'={}\{\grlabel_{\mskip-2mu t}\}{}\wedge{}\nonumber\\
&&\qquad\qquad\qquad{}\phantom{(}\theta'=\succstamp{p}(\theta)\wedge{}\aenvironment'=\aenvironment\wedge\avaluation'=\avaluation)
\renumber{\refcondition{test}[f]{\execution}}\\&&\qquad\qquad\quad\llap{${}\wedge{}$}({\statesoftrace{\tau_p}}_{k-1}=\stateofprocess{\ell}{\theta}{\aenvironment}{\avaluation}\wedge{}\nonumber\\
&&\qquad\qquad\quad\phantom{(}{\eventsoftrace{\tau_p}}_{k}=\testevent{\markertuple{p}{\ell}{ \binstra[\tagsequence]{\mneg\nonterminal{operation}}{\grlabel_{\mskip-2mu t}}}{\theta}}{}\wedge{}\nonumber\\
&&\qquad\qquad\quad\phantom{(}{\statesoftrace{\tau_p}}_{k}=\stateofprocess{\acontrol'}{\theta'}{\aenvironment'}{\avaluation'}){}\implies{}\nonumber\\
&&\qquad\qquad\qquad(\satisfiable{\evalexprenv[\nonterminal{operation}](\aenvironment,\avaluation)=0}\wedge{}\acontrol'={}\attafter(\nonterminal{b-u-instr}){}\wedge{}\nonumber\\
&&\qquad\qquad\qquad{}\phantom{(}\theta'=\succstamp{p}(\theta)\wedge{}\aenvironment'=\aenvironment\wedge\avaluation'=\avaluation)
\nonumber
\end{eqntabular}
\end{pcframed}
\caption{\LISA branches\label{fig:lisa-branches}}
\end{figure}
The
branch instruction
$\binstra[\tagsequence]{\nonterminal{operation}}{\grlabel_{\mskip-2mu t}}{}$
branches to $\grlabel_{\mskip-2mu t}$ if $\nonterminal{operation}$
is true, else continues in sequence to the next instruction. The unconditional branching
$\binstra[\tagsequence]{\true}{\grlabel_{\mskip-2mu t}}$ 
will always branch to the next label $\grlabel_{\mskip-2mu t}$. $\binstra[\tagsequence]{\false}{\grlabel_{\mskip-2mu t}}$ is equivalent
to \texttt{skip}. Branching can only be to an existing label within the same process.

\subsection{Anarchic semantics of \protect\LISA} 
The anarchic semantics of a \LISA program
$\texttt{P}$ is
\begin{eqntabular}{rcl}
\defsemantics<anarchictag>[\texttt{P}]&\triangleq&
\{\execution\in\setofallexecutions\mid \refcondition{start}{\execution} \wedge\ldots\wedge
\refcondition{test}{\execution}\}\ .\label{eq:def:anarchic-semantics}
\end{eqntabular}

\begin{example}\label{ex:LB}\normalfont
 Consider the \LISA \texttt{LB} (load buffer) program\\[0.75ex]
\mbox{}{\marge}\begin{minipage}[fl]{10cm}
\begin{verbatim}
{ x = 0; y = 0; }
P0          | P1       ;
1: r[] r1 x | 4: r[] r2 y ;
2: w[] y 1  | 5: w[] x  1 ;
3:          | 6:
exists(0:r1=1 /\ 1:r2=1)
\end{verbatim}
\end{minipage}\\[0.75ex]
\clearpage
The computational semantics $\semantics[\texttt{LB}]$ of \texttt{LB} contains the following
execution $\execution$ (stamps are useless since all events are different).
\begin{eqntabular}[fl]{@{\marge}rCl}
\execution&=&
\bullet\eventstate{w(\startprocess,\avar{x}, 0)}{\bullet}\eventstate{w(\startprocess,\avar{y}, 0)}{\bullet}\times\nonumber\\
&&\stateofprocess{\texttt{1:}}{}{\emptyset}{\emptyset}\eventstate{r(\texttt{P0},\avar{x}, \theta_{1})}{\stateofprocess{\texttt{2:}}{}{\emptyset}{\{\theta_{1}=1\}}}\;\eventstate{w(\texttt{P0},\avar{y}, 1)}{\stateofprocess{\texttt{3:}}{}{\emptyset}{\{\theta_{1}=1\}}}\times\nonumber\\
&&\stateofprocess{\texttt{4:}}{}{\emptyset}{\emptyset}\eventstate{r(\texttt{P1},\avar{y}, \theta_{4})}{\stateofprocess{\texttt{5:}}{}{\emptyset}{\{\theta_{4}=1\}}}\;\eventstate{w(\texttt{P1},\avar{x}, 1)}{\stateofprocess{\texttt{6:}}{}{\emptyset}{\{\theta_{4}=1\}}}\times\nonumber\\
&&\bullet\eventstate{r(\finishprocess,\avar{y},1)}{\bullet}\;\eventstate{r(\finishprocess,\avar{x},1)}{\bullet}\times\nonumber\\
&&\{\catrfevent{w(\texttt{P1},\avar{x}, 1)}{r(\texttt{P0},\avar{x}, \theta_{1})}),
\catrfevent{w(\texttt{P0},\avar{y}, 1)}{r(\texttt{P1},\avar{y}, \theta_{4})},\nonumber\\
&&\phantom{\{}\catrfevent{w(\texttt{P1},\avar{x}, 1)}{r(\finishprocess,\avar{x}, 1)},
\catrfevent{w(\texttt{P0},\avar{y}, 1)}{r(\finishprocess,\avar{y}, 1)}\}\renumber{\qed}
\end{eqntabular}
\end{example}
\begin{lemma}[Orderliness]\label{lem:Orderliness} The stamps of events in a trace $\tau\in\{\tau_p\mid p\in\setofallprocessidentifiers\}$ of a \LISA program are in strictly increasing order per process $p$. 
\begin{eqntabular}{@{}l@{\qquad\qquad}}
\forall p\in\setofallprocessidentifiers\suchthat\forall i,j\in\interval[open right]{1}{1+\lengthoftrace{\tau}}\suchthat{}
(\stampof(\eventsoftrace{\tau}_i)\in\setofallevents[$p$]\wedge \stampof(\eventsoftrace{\tau}_j)\in\setofallevents[$p$]\wedge i< j)\defcondition<Orderliness>{Orderliness}{\tau}
\\
\qquad{}\implies{}(\stampof(\statesoftrace{\tau}_{i+1})=\succstamp{p}(\stampof(\statesoftrace{\tau}_{i}))\wedge\stampof(\eventsoftrace{\tau}_i)\ltstamp{p}\stampof(\eventsoftrace{\tau}_j))\ .\nonumber
\end{eqntabular}
This condition \refcondition{Orderliness}{\tau} enforces \refcondition{uniqueness}{\tau}.
\qed\end{lemma}

\begin{example}\label{ex:stamps=unique-in-po}\normalfont
The choice $\setofallstamps{p} \colsep{\triangleq} \{p\}\times\setofallnaturals$, $\infstamp{p}=\pair{p}{1}$, $\succstamp{p}(\pair{p}{\theta})=\pair{p}{\theta+1}$, $\stampof(\eventsoftrace{\tau}_k)=\pair{p}{k}$, $k\in\interval[open right]{1}{1+\lengthoftrace{\tau}}$, $\pair{p'}{\theta}\ltstamp{p}\pair{p''}{\theta'}\triangleq p=p'=p''\wedge \theta<\theta'$ satisfies \refcondition{uniqueness}{\tau} and \refcondition{Orderliness}{\tau}.
\qed\end{example}

\begin{proof}[of \textsc{Lemma}~\ref{lem:Orderliness}] The events $\eventsoftrace{\tau}_i$ have the stamp 
$\stampof(\eventsoftrace{\tau}_i)$ of the state $\statesoftrace{\tau}_i$ that generate them. 
By case analysis for \LISA instructions two successive states $\statesoftrace{\tau}_i$ and $\statesoftrace{\tau}_{i+1}$ generated by an instruction of a process $p$ have stamps $\stampof(\statesoftrace{\tau}_{i+1})=\theta'=\succstamp{p}(\theta)=\stampof(\statesoftrace{\tau}_{i})$ in 
\refcondition{fence}{\execution},
\refcondition{assignment}{\execution},
\refcondition{awrite}{\execution},
\refcondition{aread}{\execution},
\refcondition{beginrmw}[b]{\execution},
\refcondition{beginrmw}[e]{\execution},
\refcondition{test}[t]{\execution},
and
\refcondition{test}[f]{\execution}.
It follows, by def.~of $\succstamp{p}$ in Sect.\nbsp\ref{sec:stamps} that $\theta\ltstamp{p}\theta'$. This extends along traces since stamps of process $p$ are not $\leqstamp{p}$-comparable with stamps of other processes. This implies \refcondition{uniqueness}{\tau} since stamps for different processes are different and for the same process are comparable since $\ltstamp{p}$ is a total order and moreover in strictly increasing order.
\end{proof}
\begin{intuition}\label{intuition:communication-determines-computation}
The following Theorem \ref{th:communication-determines-computation} shows that in an execution consisting of a computation part and a communication part,
the communication part provides enough information to rebuilt the computation part. Otherwise stated, the abstraction $\alpha_{\Gamma}(\semantics[\texttt{P}])$ $\triangleq$
$\{\rfcommunicationrelation\mid\computation\times\rfcommunicationrelation\in\semantics[\texttt{P}]\}$ is an isomorphism. Note that \LISA is deterministic but a similar result would hold with random choices. The set of executions resulting from the random choices with a given communication relation $\rfcommunicationrelation$ can be reconstructed from the communication relation $\rfcommunicationrelation$. The importance of this result is to show that to put constraints on the computations it is enough to but constraints on communications.\mbox{}\hfill\mbox{}\qed
\end{intuition}

\bgroup
\begin{graybox}
\def\thetheorem{1}
\begin{theorem}\label{th:communication-determines-computation}
In an anarchic execution $\computation \times\rfcommunicationrelation\in\semantics<anarchictag>[\texttt{P}]$, the communication $\rfcommunicationrelation$ uniquely determines the computation $\computation$.
\end{theorem}
\end{graybox}
\begin{proof}[of \textsc{Theorem}~\ref{th:communication-determines-computation}]
Let $\execution=\computation \times\rfcommunicationrelation=\parallelexecution{\trace_{\startprocess}}{p\in\setofallprocessidentifiers}{}{\trace_p}{\trace_{\finishprocess}}{\rfcommunicationrelation}\in\semantics<anarchictag>[\texttt{P}]$. ${\trace_{\startprocess}}$ depends on \texttt{P} (more precisely
$\varof\sqb{\texttt{P}}$) but not on $\rfcommunicationrelation$. Let ${p\in\setofallprocessidentifiers}$ and 
$\trace_p$ = $\langle\eventstate{\eventsoftrace{\tau_p}_{k}}{\statesoftrace{\tau_p}_{k}}\mid k\in\interval[open right]{1}{1+\lengthoftrace{\tau}}\rangle$. The proof is by induction of $k$. For $k$ = $1$, ${\eventsoftrace{\tau_p}_{k}}=\startevent$ by \refcondition{start}{\execution} and ${\statesoftrace{\tau_p}_{k}}$ is defined by \refcondition{zeroing}{\execution} so does not depend upon $\rfcommunicationrelation$. For the induction step, we have state ${\statesoftrace{\tau_p}}_{k-1}=\stateofprocess{\ell}{\theta}{\aenvironment}{\avaluation}$ and must consider all possible instructions at $\ell$ leading to the next event ${\eventsoftrace{\tau_p}}_{k}$ and state ${\statesoftrace{\tau_p}}_{k}$.
\begin{itemize}[itemsep=3pt]

\item The semantics of the fence (\refcondition{fence}{\execution}), register (\refcondition{assignment}{\execution}), write (\refcondition{awrite}{\execution}), RMW (\refcondition{beginrmw}[b]{\execution} and \refcondition{beginrmw}[e]{\execution}), and test
(\refcondition{test}[t]{\execution} and \refcondition{test}[f]{\execution}) instruction, hence ${\eventsoftrace{\tau_p}}_{k}$ and state ${\statesoftrace{\tau_p}}_{k}$, does not depend at all on the communication relation $\rfcommunicationrelation$.

\item The semantics \refcondition{aread}{\execution} of the read instruction is the only one depending on the communication relation $\rfcommunicationrelation$. The semantics \refcondition{aread}{\execution} is completetly determined by the choice of $\catrfevent{{\eventsoftrace{\tau_q}}_{j}}{{\eventsoftrace{\tau_p}}_{k}}\in\rfcommunicationrelation$.
By \refcondition{satisfaction}{\execution}, \refcondition{singleness}{\execution}, and \refcondition{match}{\execution}, this choice is unique.

\end{itemize}
It remains to consider ${\trace_{\finishprocess}}$. By \refcondition{initialisation}[f]{\execution}, ${\trace_{\finishprocess}}$ contains only read instructions, and so, by the above argument is uniquely determined by $\rfcommunicationrelation$.
\end{proof}
\egroup

\FloatBarrier
\endgroup

\newpage
\begingroup
\section{The weakly consistent semantics of \protect\LISA defined by a \protect\cat communication specification}\label{sec:cat-communication-specification-LISA}
\let\section\subsection
\let\subsection\subsubsection
\let\subsubsection\paragraph
To be language independent, the \cat communication specification \cite{AlglaveCousotMaranget-cat-2015} does rely on an abstraction of executions called \emph{candidate executions}. The abstraction essentially forget about values manipulated by programs and program instructions not related to communications. So a 
candidate execution records how communications are performed, not which values are communicated. See \cite{JA-Bertinoro} for an introduction to the \cat communication specification language and \cite{DBLP:conf/sfm/Alglave15} for models of architectures.

\section{Abstraction to a Candidate Execution}\label{sec:abstraction-to-candidate-execution}

The candidate execution abstraction
$\alphacandidateexecution\in\setofallexecutions\mapsto\setofallcandidateexecutions$
extracts a candidate execution
$\alphacandidateexecution(\execution)\in\setofallcandidateexecutions$ from an execution $\execution\in\setofallexecutions$. This candidate execution
$\alphacandidateexecution(\execution)$ is used by the \cat 
specification language semantics to decide whether that execution $\execution$ is
feasible in the weak consistency model defined by a \cat 
communication specification.

\subsection{\bfseries Events of an execution.}\label{sec:Events-of-event-trace}\quad

The candidate execution abstraction extracts the computation events of an execution.
\begin{eqntabular*}{rcl}
\defalphacandidateexecutionevt(\trace)&\triangleq&\{\epsilon\in\setofallevents\mid\exists \trace_1, \trace_2\suchthat \trace=\trace_1\: \eventstate{\epsilon}{\sigma}\: \trace_2\}\\
\alphacandidateexecutionevt({\parallelexecution{\trace_{\startprocess}}{p\in\setofallprocessidentifiers}{}{\trace_p}{\trace_{\finishprocess}}{\rfcommunicationrelation}})&\triangleq&\bigcup_{p\in\setofallprocessidentifiersstartfinish}\alphacandidateexecutionevt(\trace_p)\footnotemark
\end{eqntabular*}\footnotetext{In addition, the \sfscriptsizefont{cat} language does not allow to refer to
final reads in $\setofallreadevents({\scriptstyle\finishprocess})$.}The events $\alphacandidateexecutionevt(\execution)$ of an execution $\execution$
can be partitioned into write, read, branch, fence events, \catbeginrmw, \catendrmw, \etc.

\subsection{\bfseries Program order of an event trace.}\label{sec:Program-order}\quad
The candidate execution abstraction extracts the program order of an execution, more precisely the program execution order, \ie the pair of events generated by execution of successive actions of a process\footnote{By program order, one must understand order of execution of actions in the program, not necessarily the order in which they appear in the program text, although they are often the same. For a  counter-example of the difference between the order of actions in the program and during execution, one can imagine a silly command \texttt{execute\_next;a;b;c;} which semantics would to be to execute action \texttt{b}, then  \texttt{a}, and then \texttt{c}. So the  program syntactic order is \texttt{a}, then  \texttt{b}, and then \texttt{c} while the  program execution order is \texttt{b}, then  \texttt{a}, and then \texttt{c}.}. By convention, the
initial write events $w(\startprocess,\avar{x})$ are before any process event or final read in the program order.
\begin{eqntabular*}{rcl}
\defalphacandidateexecutionpo(\trace)
&\triangleq&
\{\pair{\epsilon}{\epsilon'}\mid\exists \trace_1, \trace_2, \trace_3\suchthat  \trace=\trace_1\: \eventstate{\epsilon}{\sigma}\: \trace_2\: \eventstate{\epsilon'}{\sigma'}\: \trace_3\}\\
\alphacandidateexecutionpo({\parallelexecution{\trace_{\startprocess}}{p\in\setofallprocessidentifiers}{}{\trace_p}{\trace_{\finishprocess}}{\rfcommunicationrelation}})
&\triangleq&
\bigcup_{{p\in\setofallprocessidentifiers}}\{\pair{\epsilon}{\epsilon'}\mid\epsilon\in\alphacandidateexecutionevt(\trace_{\startprocess})\wedge\epsilon'\in(\alphacandidateexecutionevt(\trace_p)\cup\alphacandidateexecutionevt(\trace_{\finishprocess}))\}\footnotemark{}^{,}\footnotemark{}\\
&&{}\cup{}\bigcup_{{p\in\setofallprocessidentifiers}}\alphacandidateexecutionpo(\trace_p){}\\
&&{}\cup{}\bigcup_{{p\in\setofallprocessidentifiers}}\{\pair{\epsilon}{\epsilon'}\mid\epsilon\in\alphacandidateexecutionevt(\trace_p)\wedge\epsilon'\in\alphacandidateexecutionevt(\trace_{\finishprocess})\}
\end{eqntabular*}\addtocounter{footnote}{-1}\footnotetext{The \sfscriptsizefont{herd7} tool considers the program order to be $\alphacandidateexecutionpo(\execution)\setminus(\setofallevents\times\setofallreadevents({\scriptstyle\finishprocess}))$ instead.}\addtocounter{footnote}{1}\footnotetext{The initial writes are not ordered between themselves by the program order and similarly for the final reads. This
is because if an execution of the semantics has the initial writes and final reads in some order, reshuffling them in any other order
is also a valid execution of the semantics.}\subsection{\bfseries Read-from relation.}\label{sec:Read-from-relation}\quad
The candidate execution abstraction extracts the read-from relation of an event trace modeling who reads from where.
\begin{eqntabular*}{rcl}
\defalphacandidateexecutionrf({\parallelexecution{\trace_{\startprocess}}{p\in\setofallprocessidentifiers}{}{\trace_p}{\trace_{\finishprocess}}{\rfcommunicationrelation}})&\triangleq&\rfcommunicationrelation\ .
\end{eqntabular*}
\subsection{\bfseries Initial writes.}\label{sec:Initial-writes}\quad By the \initialisation condition \refcondition{initialisation}{\semantics}, all shared variables are assumed to be initialised.
The candidate execution abstraction extracts the initial writes of an execution.

\begin{eqntabular*}{rcl}
\defalphacandidateexecutioniw({\parallelexecution{\trace_{\startprocess}}{p\in\setofallprocessidentifiers}{}{\trace_p}{\trace_{\finishprocess}}{\rfcommunicationrelation}})&\triangleq&\alphacandidateexecutionevt({\trace_{\startprocess}})\ .
\end{eqntabular*}

\subsection{\bfseries Final writes.}\label{sec:Final-writes}\quad By the \finalisation condition 
\refcondition{initialisation}[f]{\execution}
all the final values variables are assumed to be read upon program termination.
The candidate execution abstraction extracts the final writes satisfying these final reads of an event trace.
\begin{eqntabular*}{rcl}
\defalphacandidateexecutionfw({\parallelexecution{\trace_{\startprocess}}{p\in\setofallprocessidentifiers}{}{\trace_p}{\trace_{\finishprocess}}{\rfcommunicationrelation}})&\triangleq&\{w\mid\exists r\in\alphacandidateexecutionevt({\trace_{\finishprocess}})\suchthat \catrfevent{w}{r}\in\rfcommunicationrelation\}\ .
\end{eqntabular*}

\subsection{\bfseries \cat candidate executions.}\label{sec:cat-candidate-executions}\quad
The \cat candidate executions are
\begin{eqntabular}{rcl}
\setofallcandidateexecutions&\triangleq&\wp(\setofallevents)\times\wp(\setofallevents\times\setofallevents)\times \wp(\setofallwriteevents\times\setofallreadevents)\times \wp(\setofallwriteevents)\times \wp(\setofallwriteevents)
\nonumber\\[1ex]
\defalphacandidateexecution&\in&\setofallexecutions\rightarrow\setofallcandidateexecutions\nonumber\\
\alphacandidateexecution(\execution)&\triangleq&\cattuple{\alphacandidateexecutionevt(\execution)}{\alphacandidateexecutionpo(\execution)}{\alphacandidateexecutionrf(\execution)}{\alphacandidateexecutioniw(\execution)}{\alphacandidateexecutionfw(\execution)}\colsep{\in}\setofallcandidateexecutions\nonumber\\[1ex]
\defalphacandidateexecution&\in&\setofalltracesemantics\rightarrow\wp(\setofallexecutions\times\setofallcandidateexecutions)
\defdefinition{alphacandidateexecution}\\
\alphacandidateexecution(\semantics)&\triangleq&\{\pair{\execution}{\alphacandidateexecution(\execution)}\mid \execution\in\semantics\}\ .
\nonumber
\end{eqntabular}

\begin{example}\label{ex:LB-candidate-execution}\normalfont
 Continuing Ex.\ \ref{ex:LB}, we have 
\begin{eqntabular*}[fl]{rcl}
\alphacandidateexecutionevt(t)&=&
\{
w(\startprocess,\avar{x}, 0),
w(\startprocess,\avar{y}, 0),
r(\texttt{P0},\avar{x}, \theta_{1}),
w(\texttt{P0},\avar{y}, 1),
r(\texttt{P1},\avar{y}, \theta_{4}),
w(\texttt{P1},\avar{x}, 1),\\
&&\phantom{\{}r(\finishprocess,\avar{x}),
r(\finishprocess,\avar{y},1)
\}\\
\alphacandidateexecutionpo(t)&=&
\{
\pair{w(\startprocess,\avar{x}, 0)}{r(\texttt{P0},\avar{x}, \theta_{1})},
\pair{w(\startprocess,\avar{x}, 0)}{w(\texttt{P0},\avar{y}, 1)},
\pair{w(\startprocess,\avar{x}, 0)}{r(\texttt{P1},\avar{y}, \theta_{4})},\\&&\phantom{\{}
\pair{w(\startprocess,\avar{x}, 0)}{w(\texttt{P1},\avar{x}, 1)},
\pair{w(\startprocess,\avar{x}, 0)}{r(\finishprocess,\avar{x})},
\pair{w(\startprocess,\avar{x}, 0)}{r(\finishprocess,\avar{y},1)},\\&&\phantom{\{}
\pair{w(\startprocess,\avar{y}, 0)}{r(\texttt{P0},\avar{x}, \theta_{1})},
\pair{w(\startprocess,\avar{y}, 0)}{w(\texttt{P0},\avar{y}, 1)},
\pair{w(\startprocess,\avar{y}, 0)}{r(\texttt{P1},\avar{y}, \theta_{4})},\\&&\phantom{\{}
\pair{w(\startprocess,\avar{y}, 0)}{w(\texttt{P1},\avar{x}, 1)},
\pair{w(\startprocess,\avar{y}, 0)}{r(\finishprocess,\avar{x})},
\pair{w(\startprocess,\avar{y}, 0)}{r(\finishprocess,\avar{y},1)},\\&&\phantom{\{}
\pair{r(\texttt{P0},\avar{x}, \theta_{1})}{w(\texttt{P0},\avar{y}, 1)},
\pair{r(\texttt{P0},\avar{x}, \theta_{1})}{r(\texttt{P1},\avar{y}, \theta_{4})},
\pair{r(\texttt{P0},\avar{x}, \theta_{1})}{w(\texttt{P1},\avar{x}, 1)},\\&&\phantom{\{}
\pair{r(\texttt{P0},\avar{x}, \theta_{1})}{r(\finishprocess,\avar{x})},
\pair{r(\texttt{P0},\avar{x}, \theta_{1})}{r(\finishprocess,\avar{y},1)},
\pair{w(\texttt{P0},\avar{y}, 1)}{r(\texttt{P1},\avar{y}, \theta_{4})},\\&&\phantom{\{}
\pair{w(\texttt{P0},\avar{y}, 1)}{w(\texttt{P1},\avar{x}, 1)},
\pair{w(\texttt{P0},\avar{y}, 1)}{r(\finishprocess,\avar{x})},
\pair{w(\texttt{P0},\avar{y}, 1)}{r(\finishprocess,\avar{y},1)},\\&&\phantom{\{}
\pair{r(\texttt{P1},\avar{y}, \theta_{4})}{w(\texttt{P1},\avar{x}, 1)},
\pair{r(\texttt{P1},\avar{y}, \theta_{4})}{r(\finishprocess,\avar{x})},
\pair{r(\texttt{P1},\avar{y}, \theta_{4})}{r(\finishprocess,\avar{y},1)},\\&&\phantom{\{}
\pair{w(\texttt{P1},\avar{x}, 1)}{r(\finishprocess,\avar{x})},
\pair{w(\texttt{P1},\avar{x}, 1)}{r(\finishprocess,\avar{y},1)}
\}\\
\alphacandidateexecutionrf(t)&=&
\{
\pair{w(\texttt{P1},\avar{x}, 1)}{r(\texttt{P0},\avar{x}, \theta_{1})}),
\pair{w(\texttt{P0},\avar{y}, 1)}{r(\texttt{P1},\avar{y}, \theta_{4})},
\pair{w(\texttt{P1},\avar{x}, 1)}{r(\finishprocess,\avar{x}, 1)},\\&&\phantom{\{}
\pair{w(\texttt{P0},\avar{y}, 1)}{r(\finishprocess,\avar{y}, 1)}
\}\\
\alphacandidateexecutioniw(t)&=&
\{
w(\startprocess,\avar{x}, 0),
w(\startprocess,\avar{y}, 0)
\}\\
\alphacandidateexecutionfw(t)&=&
\{
{w(\texttt{P1},\avar{x}, 1)},
{w(\texttt{P0},\avar{y}, 1)}
\}\ .
\end{eqntabular*}
This is a non-SC candidate execution because of its cycle in union of program order and communications \cite{JA-Bertinoro}: 
\begin{eqntabular*}{C}
\includegraphics[scale=0.5,bb=0 0 293 160]{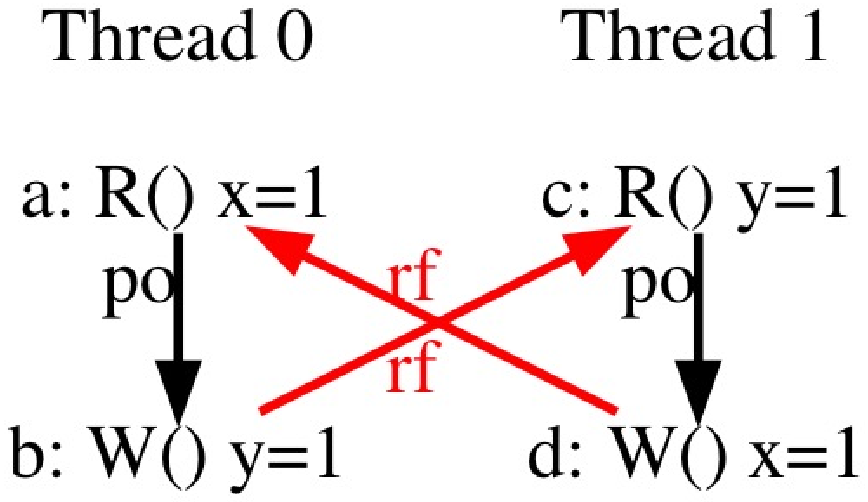}
\end{eqntabular*}
which would be invalid with the following \texttt{cat} specification\\
\begin{eqntabular}[fl]{@{\marge}L}
\texttt{acyclic (po | rf)+}\renumber{\qed}
\end{eqntabular}
\end{example}

\section{Abstraction to a semantics with weak consistency model}\label{sec:abstraction-to-semantics-with-WCM}

\subsection{\bfseries The semantics of a \cat weak consistency model specification.}\label{sec:cat-specification}\quad

\newcommand{\catallowed}{\texttt{allowed}}
\newcommand{\catforbidden}{\texttt{forbidden}}
\makeatletter
\newcommand{\anarchicsemantics}{\semantics<anarchictag>}
\makeatother
\newcommand{\catflagdomain}{\mathcal{F}}

The semantics $\catsemantics[\Hcom]\:\candidateexecution$ of a candidate execution $\candidateexecution=\pair{\computation}{\texttt{rf}}\in\setofallcandidateexecutions$ defined in \cite{AlglaveCousotMaranget-cat-HSA-2015} returns a set of answers of the form 
$\triple{j}{f}{\communicationrelation}$ where $j=\{\catallowed,\catforbidden\}$, $f\in\catflagdomain$ is the set of flags that have been set up on $\candidateexecution$ and $\communicationrelation$, and $\communicationrelation$ defines the communication relation for the execution to be \catallowed/\catforbidden. This is extended to a set $\setofcandidateexecutions\in\wp(\setofallcandidateexecutions)$ of candidate executions as
\begin{eqntabular*}{rcl}
\defalphacandidateexecution(\setofcandidateexecutions)
&\triangleq&
\{\pair{\candidateexecution}{\alphacandidateexecution(\candidateexecution)}\mid
\candidateexecution\in\setofcandidateexecutions\}
\\
\end{eqntabular*}

\subsection{\bfseries Computational semantics with weak consistency model.}\label{sec:cat-abstraction}\quad
 The computational semantics $\semantics$ restricted by a weak consistency model specified by \cat specification $\Hcom$ is then $\semantics$ $\triangleq$
$\alphacat[\Hcom]\comp\alphacandidateexecution(\anarchicsemantics[\texttt{P}])$ where
\begin{eqntabular}{rcl}
\defalphacat[\Hcom](\mathcal{C})
&\triangleq&
\{\triple{\computation}{\texttt{rf}}{\communicationrelation}\mid
\begin{array}[t]{@{}l@{}}
\pair{\pair{\computation}{\texttt{rf}}}{\candidateexecution}\in\mathcal{C}
\wedge{}\\[0.33ex]
\quad\exists f\in\catflagdomain\suchthat
\triple{ \catallowed}{f}{\communicationrelation}\in\catsemantics[\Hcom]\:\candidateexecution\}
\end{array}
\defdefinition{alphacat-event-semantics}
\end{eqntabular}
\endgroup


\begin{thebibliography}{14}
\providecommand{\natexlab}[1]{#1}
\providecommand{\url}[1]{\texttt{#1}}
\expandafter\ifx\csname urlstyle\endcsname\relax
  \providecommand{\doi}[1]{doi: #1}\else
  \providecommand{\doi}{doi: \begingroup \urlstyle{rm}\Url}\fi

\bibitem[Alglave(2015{\natexlab{a}})]{DBLP:conf/sfm/Alglave15}
Jade Alglave.
\newblock Modeling of architectures.
\newblock In Marco Bernardo and Einar~Broch Johnsen, editors, \emph{Formal
  Methods for Multicore Programming - 15th International School on Formal
  Methods for the Design of Computer, Communication, and Software Systems,
  {SFM} 2015, Bertinoro, Italy, June 15-19, 2015, Advanced Lectures}, volume
  9104 of \emph{Lecture Notes in Computer Science}, pages 97--145. Springer,
  2015{\natexlab{a}}.
\newblock ISBN 978-3-319-18940-6.
\newblock \doi{10.1007/978-3-319-18941-3_3}.
\newblock URL \url{http://dx.doi.org/10.1007/978-3-319-18941-3_3}.

\bibitem[Alglave(2015{\natexlab{b}})]{JA-Bertinoro}
Jade Alglave.
\newblock I can't dance: adventures in herding cats.
\newblock Lecture notes for Bertorino summer school, March 2015{\natexlab{b}}.

\bibitem[Alglave and Maranget(2015)]{herd7}
Jade Alglave and Luc Maranget.
\newblock \textsf{\fontsize{7}{10}\selectfont herd7}.
\newblock
  \href{http://virginia.cs.ucl.ac.uk/herd}{\url{virginia.cs.ucl.ac.uk/herd}},
  31 August 2015.

\bibitem[Alglave et~al.(2014)Alglave, Maranget, and
  Tautschnig]{DBLP:journals/toplas/AlglaveMT14}
Jade Alglave, Luc Maranget, and Michael Tautschnig.
\newblock Herding cats: Modelling, simulation, testing, and data mining for
  weak memory.
\newblock \emph{{ACM} Trans. Program. Lang. Syst.}, 36\penalty0 (2):\penalty0
  7:1--7:74, 2014.
\newblock \doi{10.1145/2627752}.
\newblock URL \url{http://doi.acm.org/10.1145/2627752}.

\bibitem[Alglave et~al.(2015{\natexlab{a}})Alglave, Batty, Donaldson,
  Gopalakrishnan, Ketema, Poetzl, Sorensen, and Wickerson]{abd15}
Jade Alglave, Mark Batty, Alastair~F. Donaldson, Ganesh Gopalakrishnan, Jeroen
  Ketema, Daniel Poetzl, Tyler Sorensen, and John Wickerson.
\newblock {GPU} concurrency: Weak behaviours and programming assumptions.
\newblock In \emph{ASPLOS}, 2015{\natexlab{a}}.

\bibitem[Alglave et~al.(2015{\natexlab{b}})Alglave, Cousot, and
  Maranget]{AlglaveCousotMaranget-cat-2015}
Jade Alglave, Patrick Cousot, and Luc Maranget.
\newblock La langue au chat: \ttfont{cat}, a language to describe consistency
  properties.
\newblock Unpublished manuscript, 31 January 2015{\natexlab{b}}.

\bibitem[Alglave et~al.(2015{\natexlab{c}})Alglave, Cousot, and
  Maranget]{AlglaveCousotMaranget-cat-HSA-2015}
Jade Alglave, Patrick Cousot, and Luc Maranget.
\newblock Syntax and semantics of the \cat language.
\newblock \emph{HSA Foundation}, Version 1.1:\penalty0 38 p., 16 Oct
  2015{\natexlab{c}}.
\newblock URL \url{http://www.hsafoundation.com/?ddownload=5382}.

\bibitem[Hennessy and Plotkin(1979)]{DBLP:conf/mfcs/HennessyP79}
Matthew Hennessy and Gordon~D. Plotkin.
\newblock Full abstraction for a simple parallel programming language.
\newblock In Jir{\'{\i}} Becv{\'{a}}r, editor, \emph{Mathematical Foundations
  of Computer Science 1979, Proceedings, 8th Symposium, Olomouc,
  Czechoslovakia, September 3-7, 1979}, volume~74 of \emph{Lecture Notes in
  Computer Science}, pages 108--120. Springer, 1979.
\newblock \doi{10.1007/3-540-09526-8_8}.
\newblock URL \url{http://dx.doi.org/10.1007/3-540-09526-8_8}.

\bibitem[{HSA~Foundation}(2015)]{HSA-Foundation-PSAS-2015}
{HSA~Foundation}.
\newblock Hsa platform system architecture specification 1.0.
\newblock
  \href{http://www.hsafoundation.com/?ddownload=4944}{\texttt{HSA-SysArch-1.01.pdf}},
  \href{http://www.hsafoundation.com/?ddownload=5381}{\texttt{cat\_ModelExpressions-1.1.pdf}},
  15 January 2015.

\bibitem[Keller(1976)]{DBLP:journals/cacm/Keller76}
Robert~M. Keller.
\newblock Formal verification of parallel programs.
\newblock \emph{Commun. {ACM}}, 19\penalty0 (7):\penalty0 371--384, 1976.
\newblock \doi{10.1145/360248.360251}.
\newblock URL \url{http://doi.acm.org/10.1145/360248.360251}.

\bibitem[King(1976)]{DBLP:journals/cacm/King76}
James~C. King.
\newblock Symbolic execution and program testing.
\newblock \emph{Commun. {ACM}}, 19\penalty0 (7):\penalty0 385--394, 1976.
\newblock \doi{10.1145/360248.360252}.
\newblock URL \url{http://doi.acm.org/10.1145/360248.360252}.

\bibitem[Knuth(1990)]{DBLP:conf/waga/Knuth90}
Donald~E. Knuth.
\newblock The genesis of attribute grammars.
\newblock In Pierre Deransart and Martin Jourdan, editors, \emph{Attribute
  Grammars and their Applications, International Conference WAGA, Paris,
  France, September 19-21, 1990, Proceedings}, volume 461 of \emph{Lecture
  Notes in Computer Science}, pages 1--12. Springer, 1990.
\newblock ISBN 3-540-53101-7.
\newblock \doi{10.1007/3-540-53101-7_1}.
\newblock URL \url{http://dx.doi.org/10.1007/3-540-53101-7_1}.

\bibitem[Lamport(1979)]{DBLP:journals/tc/Lamport79}
Leslie Lamport.
\newblock How to make a multiprocessor computer that correctly executes
  multiprocess programs.
\newblock \emph{{IEEE} Trans. Computers}, 28\penalty0 (9):\penalty0 690--691,
  1979.
\newblock \doi{10.1109/TC.1979.1675439}.
\newblock URL \url{http://dx.doi.org/10.1109/TC.1979.1675439}.

\bibitem[Paakki(1995)]{DBLP:journals/csur/Paakki95}
Jukka Paakki.
\newblock Attribute grammar paradigms - {A} high-level methodology in language
  implementation.
\newblock \emph{{ACM} Comput. Surv.}, 27\penalty0 (2):\penalty0 196--255, 1995.
\newblock \doi{10.1145/210376.197409}.
\newblock URL \url{http://doi.acm.org/10.1145/210376.197409}.

\end{thebibliography}
\end{document}